\documentclass[english,twocolumn,8pt]{article}

\usepackage[utf8]{inputenc}
\usepackage[T1]{fontenc}

\usepackage{import, xspace, url, breakurl, capt-of, titling, abstract, mathtools}
\usepackage{amsfonts,braket,times,graphicx,color,siunitx,enumitem,authblk,lmodern,qcircuit}
\graphicspath{{figures/}}

\usepackage[labelfont=bf]{caption}
\usepackage[caption=false]{subfig}
\usepackage{textcomp}
\usepackage[section]{placeins}
\usepackage[breaklinks]{hyperref}
\usepackage[dvipsnames]{xcolor}
\usepackage{pdfpages}

\renewcommand{\figurename}{Fig.}

\DeclareCaptionLabelSeparator{bar}{ | }
\title{Universal control of a six-qubit quantum processor in silicon}
\author[1]{ Stephan G.J. Philips$^*$}
\author[1]{ Mateusz T. M\k{a}dzik\thanks{These authors contributed equally}}
\author[1]{ Sergey V. Amitonov}
\author[1]{ Sander L. de Snoo}
\author[1]{ Maximilian Russ}
\author[1]{ Nima Kalhor}
\author[1]{ Christian Volk} 
\author[1]{ William I.L. Lawrie} 
\author[2]{ Delphine Brousse}
\author[2]{ Larysa Tryputen}
\author[1]{ Brian Paquelet Wuetz}
\author[2]{ Amir Sammak}
\author[1]{ Menno Veldhorst}
\author[1]{ Giordano Scappucci}
\author[1]{ Lieven M.K. Vandersypen\thanks{To whom correspondence should be addressed; E-mail: L.M.K.Vandersypen@tudelft.nl}
}

\affil[1]{QuTech and the Kavli Institute of Nanoscience, Delft University of Technology, 2600 GA Delft, The Netherlands.}
\affil[2]{QuTech and Netherlands Organization for Applied Scientific Research (TNO), Delft, The Netherlands.}

\begin{document} 
\twocolumn[
  \maketitle             
  \textbf{Future quantum computers capable of solving relevant problems will require a large number of qubits that can be operated reliably. However, the requirements of having a large qubit count and operating with high-fidelity are typically conflicting. Spins in semiconductor quantum dots show long-term promise but demonstrations so far use between one and four qubits and typically optimize the fidelity of either single- or two-qubit operations, or initialization and readout. Here we increase the number of qubits and simultaneously achieve respectable fidelities for universal operation, state preparation and measurement. We design, fabricate and operate a six-qubit processor with a focus on careful Hamiltonian engineering, on a high level of abstraction to program the quantum circuits and on efficient background calibration, all of which are essential to achieve high fidelities on this extended system. State preparation combines initialization by measurement and real-time feedback with quantum-non-demolition measurements. These advances will allow for testing of increasingly meaningful quantum protocols and constitute a major stepping stone towards large-scale quantum computers. }
\\]
\saythanks
On the path to practical large-scale quantum computation, electron spin qubits in semiconductor quantum dots~\cite{vandersypenPhysToday} show promise due to their inherent potential for scaling through their small size \cite{borselli2011pauli,zajac2015reconfigurable}, long-lived coherence~\cite{veldhorst2014addressable} and compatibility with advanced semiconductor manufacturing techniques~\cite{zwerver2021qubits}. Nevertheless, spin qubits currently lag behind in scale when compared to superconducting, trapped ions and photonic platforms, which have demonstrated control of several dozen qubits~\cite{arute2019quantum, egan2021fault, zhong2020quantum}. By comparison, using semiconductor spin qubits, control of up to four qubits was achieved~\cite{hendrickx2021four} and entanglement of up to three qubits was quantified~\cite{takeda2021quantum, mkadzik2021precision, takeda2022quantum}.

Furthermore, the experience with other qubit platforms shows that in scaling up, maintaining the quality of the control requires significant efforts, for instance to deal with the denser motional spectrum in trapped ions~\cite{bruzewicz2019trapped}, to avert cross-talk in superconducting circuits~\cite{zhang2020high} or to avoid increased losses in photonic circuits~\cite{arrazola2021quantum}.  For small semiconductor spin qubit systems, state-of-the-art single-qubit gate fidelities exceed 99.9\%~\cite{yoneda2018quantum, yang2019silicon, lawrie2021simultaneous} and two-qubit gates well above 99\% fidelity have been demonstrated recently~\cite{xue2021computing,noiri2021fast,mills2021two, mkadzik2021precision}. Most quantum dot based demonstrations suffer from rather low initialization or readout fidelities, with typical visibilities of no more than 60-75\%, with one recent exception~\cite{mills2021two}. Conversely, high-fidelity spin readout has been claimed based on an analysis of the readout error mechanisms, but these claims have not been validated in combination with high-fidelity qubit control~\cite{connors2020rapid, harvey2018high}. While high-fidelity initialization, readout, single-qubit gates and two-qubit gates have thus been demonstrated individually in small systems, almost invariably one or more of these parameters are significantly comprised while optimizing others. A major challenge and important direction for the field is therefore to achieve high fidelities for all components while at the same time enlarging the qubit count.

Here we study a system of six spin qubits in a linear quantum dot array and test what performance can be achieved using known methods such as multi-layer gate patterns for independent control of the two-qubit exchange interaction~\cite{angus2007gate,zajac2016scalable, lawrie_quantum_2020} and micromagnet gradients for electric-dipole spin resonance and selective qubit addressing~\cite{pioro2007micromagnets}. Furthermore, we introduce several novel techniques for semiconductor qubits that, collectively, are critical to improve on the results and facilitate scalability, such as initialization by measurement using real-time feedback~\cite{riste2012initialization}, qubit initialization and measurement without reservoir access, and efficient calibration routines. Initialization and readout circuits span over the full six-qubit array. We characterize the quality of the control by preparing maximally entangled states of two and three spins across the array.

\begin{figure*}[!ht]
    \centering
    \includegraphics[width=\textwidth]{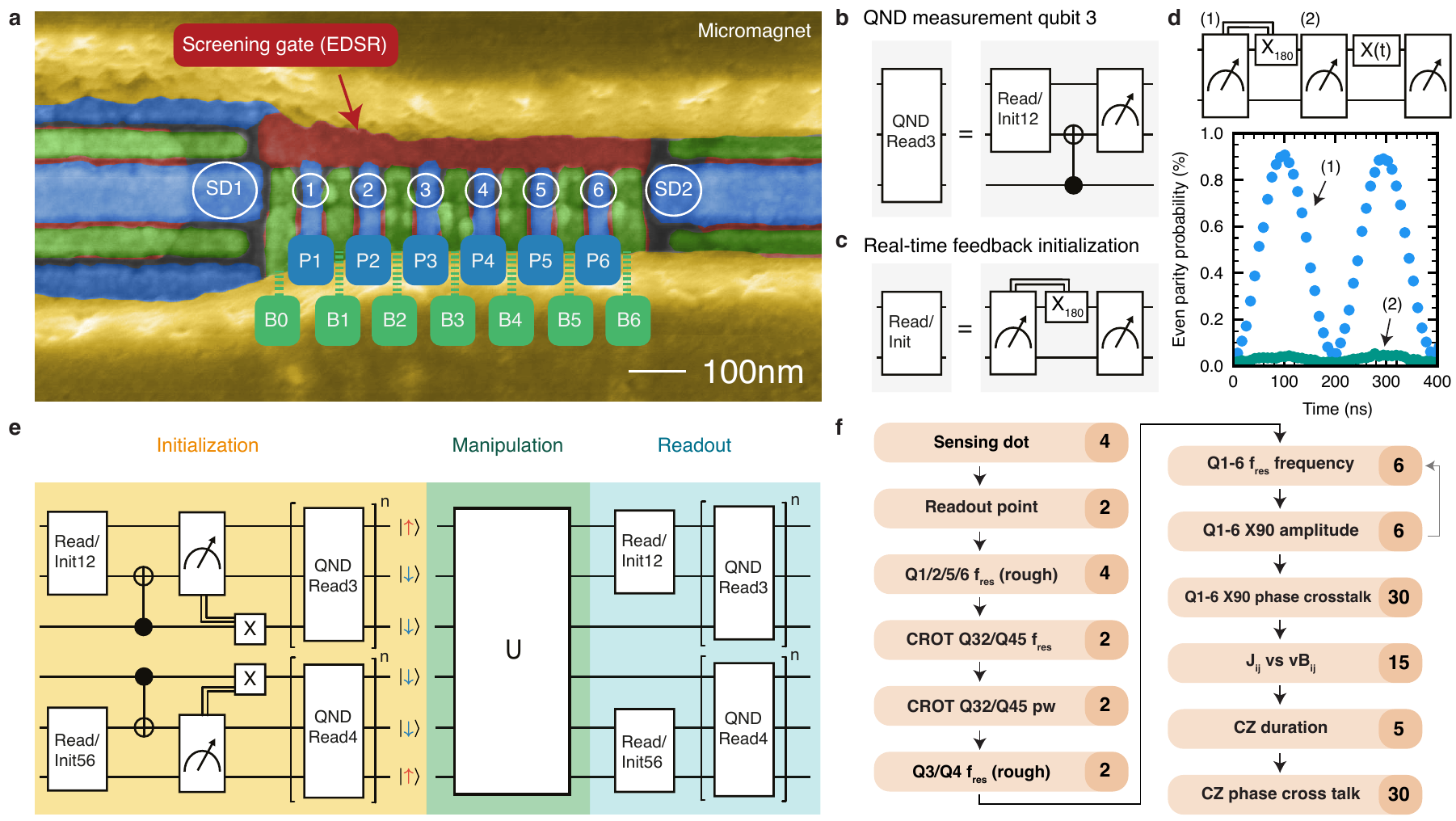}
    \caption{\textbf{Device initialization, measurement and calibration} \textbf{a}, A false-colored scanning electron microscope image of a device similar to the one used in the experiments. Each color represents a different metallization layer. Plunger (blue) and barrier (green) gates are used to define quantum dots in the channel between the screening gates (red). Two cobalt micromagnets (yellow) are placed on top of the gate stack. \textbf{b-c}, Buildings blocks used for readout and initialization in this experiment. Panel \textbf{b} shows the circuit used to perform a single QND measurement of qubit 3, panel \textbf{c} shows the circuit used for spin measurement and initialization using a parity measurement. The double line in the diagram indicates that  X$_{180}$ rotation is conditional on the measurement outcome. \textbf{d}, shows an example of a conditional rotation used to initialize the qubits. The sequence shown is applied repeatedly with short time intervals, with the final state of one cycle being the initial state of the next (up to spin relaxation). (1) shows the even parity probability of the first measurement, (2) shows the even parity probability after the bit flip conditional on the first measurement outcome. \textbf{e}, Schematic showing the total scheme used for the initialization and readout of all six qubits. \textbf{f}, Calibration graph used in the experiments. The numbers on the right show the number of parameters that are calibrated in each step.}
    \label{fig:main_fig_0}
\end{figure*}

The six-qubit array is defined electrostatically in the $^{28}$Si quantum well of a $^{28}$Si/SiGe heterostructure, between two sensing quantum dots, as seen in Figure~\ref{fig:main_fig_0}a (see Methods). The multi-layer gate pattern allows for excellent control of the charge occupation of each quantum dot, and of the tunnel couplings between neighbouring quantum dots. These parameters are controlled independently through linear combinations of gate voltages, known as virtual gates~\cite{volk2019loading}. The inter-dot pitch is chosen to be 90~nm, which for this 30 nm deep quantum well yields easy access to the regime with one electron in each dot, for short indicated as the (1,1,1,1,1,1) charge occupation. Low valley splittings on Si/SiGe devices have hindered progress in the past~\cite{kawakami2014electrical}, but in this device all valley splittings are in the range of 100-300~\textmu eV (See Supplementary materials).

\begin{figure*}[!h]
    \centering
\includegraphics[width=\textwidth]{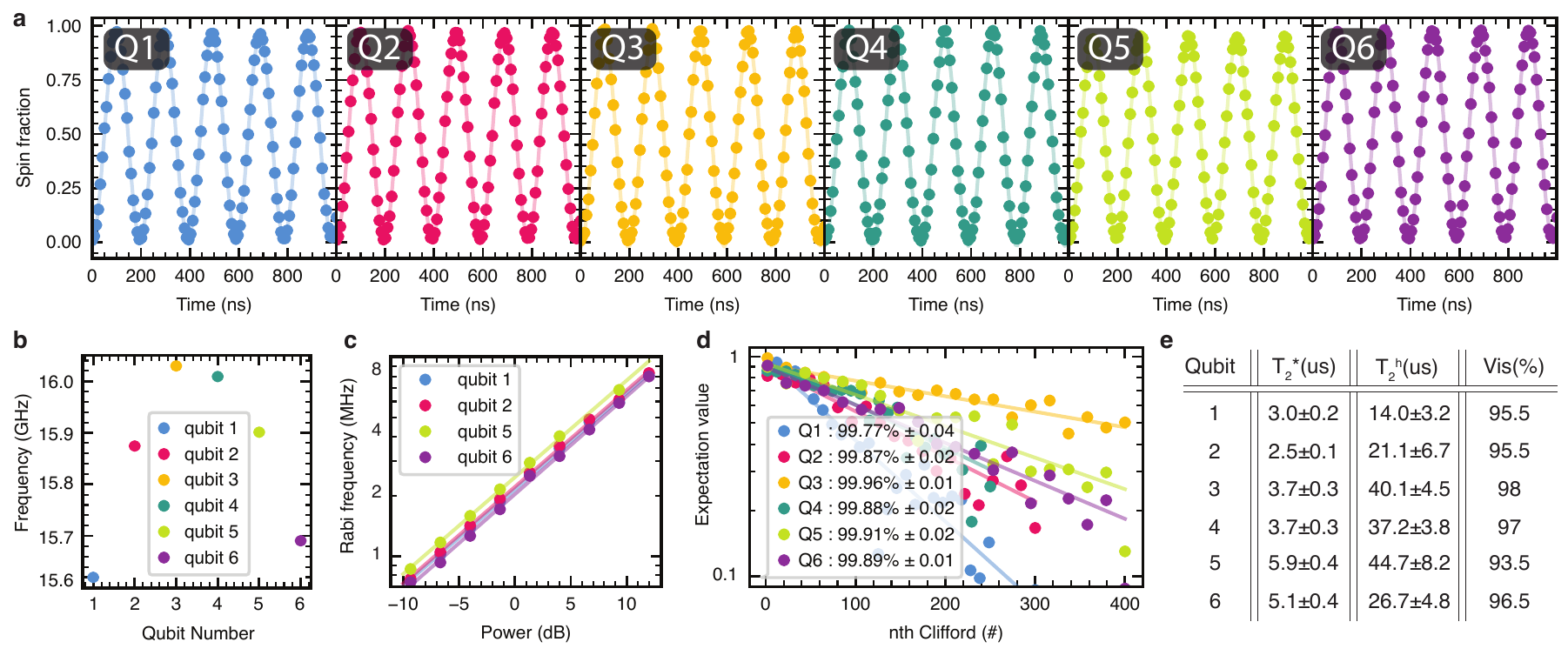}
    \caption{\textbf{Experiment overview}
    \textbf{a}, Rabi oscillations for every qubit, taken sequentially. The spin fraction refers to the spin-up fraction for qubits 2-5 and to the spin-down fraction for qubits 1 and 6. The drive amplitudes were adjusted in order to obtain uniform Rabi frequencies of 5 MHz. \textbf{b}, Qubit frequency for each the six qubits. \textbf{c}, Rabi frequency of each qubit as a function of the applied microwave power. \textbf{d}, Randomized benchmarking results for each qubit, using a 5 MHz Rabi frequency. The reported fidelity is the average single-qubit gate fidelity. The error bars ($2\sigma$) are calculated using the covariance matrix of the fit. \textbf{e,} Table showing the dephasing time T$_2^*$, Hahn echo decay time and visibilities for each qubit.}
    \label{fig:main_fig_1}
\end{figure*}

In designing the qubit measurement scheme, we focused on achieving short measurement cycles in combination with high-fidelity readout, as this accelerates testing of all other aspects of the experiment. For measuring the outer qubit pairs, we use Pauli-Spin blockade (PSB) to probe the parity of the two spins (rather than differentiating between singlet and triplet states), exploiting the fact that the T$_0$ triplet relaxes to the singlet well before the end of the 10~\textmu s readout window. We tune the outer dot pairs of the array to the (3,1) electron occupation, where the readout window is larger than in the (1,1) regime (see Extended Data \ref{ED:parity}). Since the sensing dots are less sensitive to the charge transition between the center dots, the middle qubits are measured by quantum-non-demolition (QND) measurements that map the state of qubit 3(4) on qubit 2(5) via a conditional rotation (CROT) (Fig. \ref{fig:main_fig_0}b) \cite{yoneda2018quantum, xue2020repetitive}. In this way, for every iteration of the experiment, 4 bits of information are retrieved which depend on the state of all 6 physical qubits. Iterative operation permits full readout of the 6-qubit system.

Qubit initialization is based on measurements of the spin state across the array followed by real-time feedback to place all qubits in the target initial state. This scheme has the benefit of not relying on slow thermalization and that no access to electron reservoirs is needed to bring in fresh electrons, which is helpful for scaling to larger arrays. In fact, we had experiment runs of more than one month in which the electrons stayed within the array continuously. For qubits 3 and 4, real-time feedback simply consists of flipping the qubit if the measurement returned $\ket{\uparrow}$. Initialization of qubits 1-2 (or 5-6) using parity measurements and real-time feedback is illustrated in Figure~\ref{fig:main_fig_0}d. First, assuming that the qubits start from a random state, we perform a parity measurement that will cause the state to either collapse to an even ($\ket{\downarrow \downarrow}$, $\ket{\uparrow \uparrow}$) or odd ($\ket{\uparrow \downarrow}$/$\ket{\downarrow\uparrow}$) parity (see Methods). After the measurement, a $\pi$ pulse is applied to qubit 1 in case of even parity, which converts the state to odd parity (feedback latency 660~ns). Subsequently, we perform a second measurement, which converts either of the odd parity states to $\ket{\uparrow \downarrow}$. Specifically, when pulsing towards the readout operating point, both $\ket{\uparrow \downarrow} $ and $\ket{\downarrow\uparrow}$ relax into the singlet state ((4,0) charge occupation). When pulsing adiabatically from the (4,0) back to the (3,1) charge configuration, the singlet is mapped onto the $\ket{\uparrow \downarrow}$ state. If the qubit initialization is successful, the second measurement should return an odd parity (typically $\sim$95\% success rate). To further boost the initialization fidelity we use the outcome of the second measurement to post-select successful experiment runs (see Extended Data Fig. \ref{ED:parity}d). Figure~\ref{fig:main_fig_0}d shows initialization by measurement of the first two qubits. The first readout outcome (blue) shows Rabi oscillations controlled by a microwave burst of variable duration applied near the end of the previous cycle (see methods for more details). The second readout outcome (green) shows the state after the real-time classical feedback step. The oscillation has largely vanished, indicating successful initialization by measurement and feedback.

The sequence to initialize and measure all qubits is shown in Figure~\ref{fig:main_fig_0}e (see Extended Data \ref{ED:init_seq} for the unfolded quantum circuit). We sequentially initialize qubit pair 5-6, then qubit 4, then qubits 1-2, and finally qubit 3, using the steps described above (for compactness, the steps appear as simultaneous in the diagram). In order to further enhance the measurement and initialization fidelities, we repeat the QND measurement three times, alternating the order of qubit 3 and 4 measurements. We post-select runs with three identical QND readout outcomes in both the initialization and measurement steps (except for Fig. 5 below, where readout simply uses majority voting). After performing the full initialization procedure depicted in Fig.~\ref{fig:main_fig_0}e, the six qubit array is initialized in the state $\ket{\uparrow \downarrow \downarrow \downarrow \downarrow \uparrow}$. In all measurements below, we initialize either two, three or all six qubits, depending on the requirement of the specific quantum circuit we intend to run. We leave the unused qubits randomly initialized, as the visibilities decrease when initializing all 6 qubits within a single shot sequence (see Extended Data \ref{ED:visibility}). When operating on individual qubits, the initialization and measurement procedures yield visibilities of 93.5-98\% (see Fig.~\ref{fig:main_fig_1}e). To put these numbers in perspective, if the readout error for both $\ket{0}$ and $\ket{1}$ were 1\% alongside an initialization error of 1\%, the visibility would be 96\%.

\begin{figure}[]
    \includegraphics{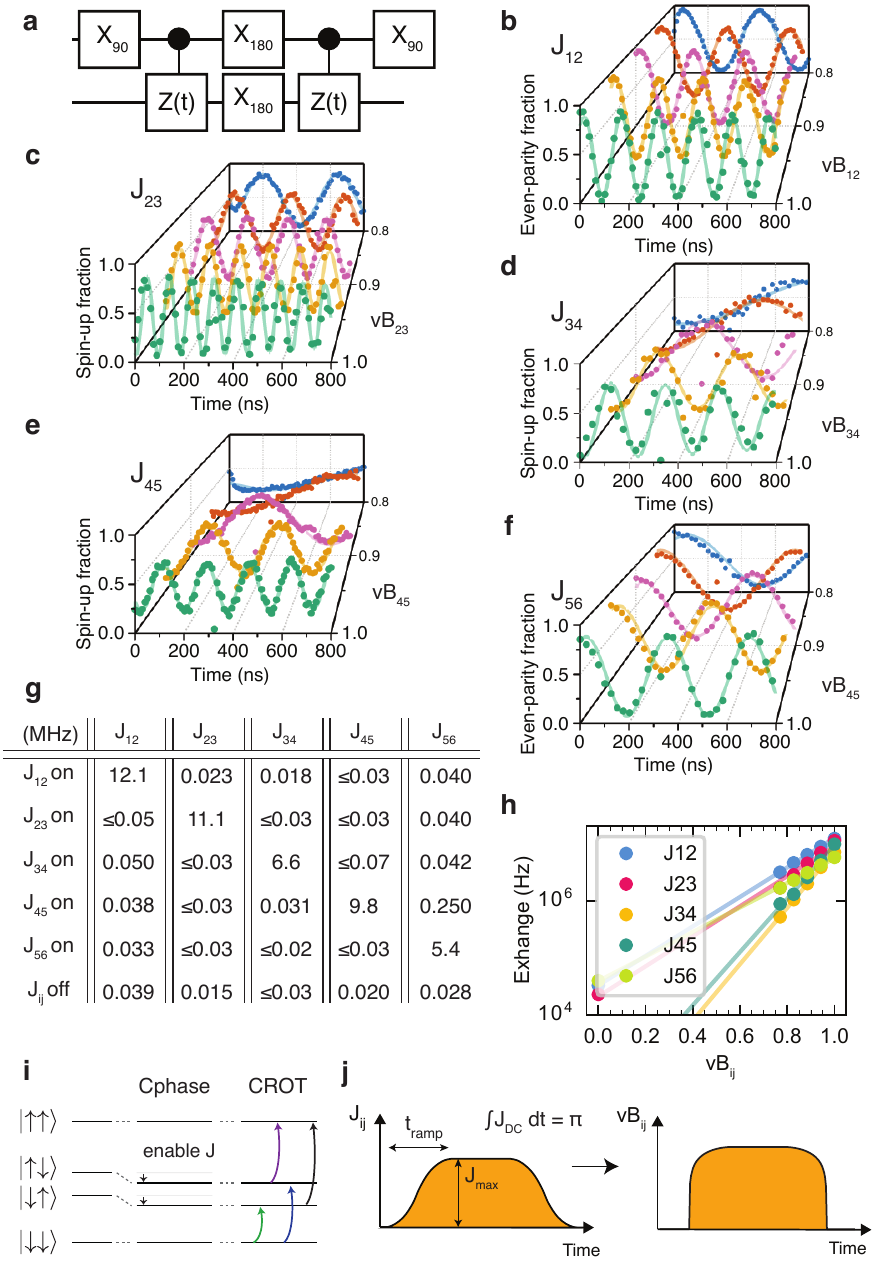}
    \caption{\textbf{Two-qubit gate characterization}
    \textbf{a}, Quantum circuit used to measure CPhase oscillations between a pair of qubits. \textbf{b-f}, Measured spin probabilities as a function of the total evolution time $2t$ for neighbouring qubit pairs for different virtual barrier gate voltages (with 0 and 1 corresponding to the exchange switched off and at its maximum value). \textbf{g}, Maximum exchange coupling measured for each qubit pair, and the corresponding residual exchange coupling for the other pairs, achievable within the AWG pulsing range without DC retuning. Bottom row: $J_{\mathrm{ij}}$ with all exchange couplings switched off (see Supplementary Data for error bars). \textbf{h}, Exchange coupling versus virtual barrier gate voltage for all qubit pairs. \textbf{i}, Schematic showing the energy levels in the absence (left) and presence (middle, right) of the effective Ising ZZ interaction under exchange (see main text). Due to the ZZ coupling term, the antiparallel spin states are lowered in energy, and pick up an additional phase as a function of time, resulting in a CPhase evolution. The shifted energy levels also allow for conditional microwave-driven rotations (CROT), which we use during initialization and readout. \textbf{j}, Pulse shape of the exchange amplitude throughout a gate voltage pulse used for the CZ gate, and the corresponding pulse shape converted to gate voltage.}
    \label{fig:main_fig_2}
\end{figure}

We manipulate the qubits via electric-dipole spin resonance (EDSR) \cite{obata2010coherent}. A micromagnet located above the gate-stack is designed to provide both qubit addressability and a driving field gradient (see Fig.~\ref{fig:main_fig_0}a and Supplementary Data). We can address each qubit individually and drive coherent Rabi oscillations as depicted in figure \ref{fig:main_fig_1}. We observe no visible damping in the first five periods. The data in figure \ref{fig:main_fig_1}c shows that the qubit frequencies are not spaced linearly, deviating from our prediction based on numerical simulations of the magnetic field gradients (Supplementary Figure 1). However, the smallest qubit frequency separation of $\approx$20~MHz is sufficient for selective qubit addressing with our operating speeds varying between 2~MHz and 5~MHz. The Rabi frequency is linear in the driving amplitude over the typical range of microwave power used in the experiment (Fig.~\ref{fig:main_fig_1}c). We operate single-qubit gates sequentially, to ensure we stay in this linear regime and to keep the calibration simple. We characterize the single-qubit properties of each qubit separately. Figure \ref{fig:main_fig_1}d shows results of randomized benchmarking experiments. All average single-qubit gate fidelities are between 99.77\% $\pm$ 0.04 and 99.96\% $\pm$ 0.01, which demonstrates that even within this extended qubit array, we retain  high-fidelity single-qubit control. The coherence times of each qubit are tabulated in figure \ref{fig:main_fig_1}e. We expect spin coherence to be limited by charge noise coupling in by the micromagnet \cite{kha2015micromagnets}.

\begin{figure*}[!h]
	\centering
	\includegraphics[width=\textwidth]{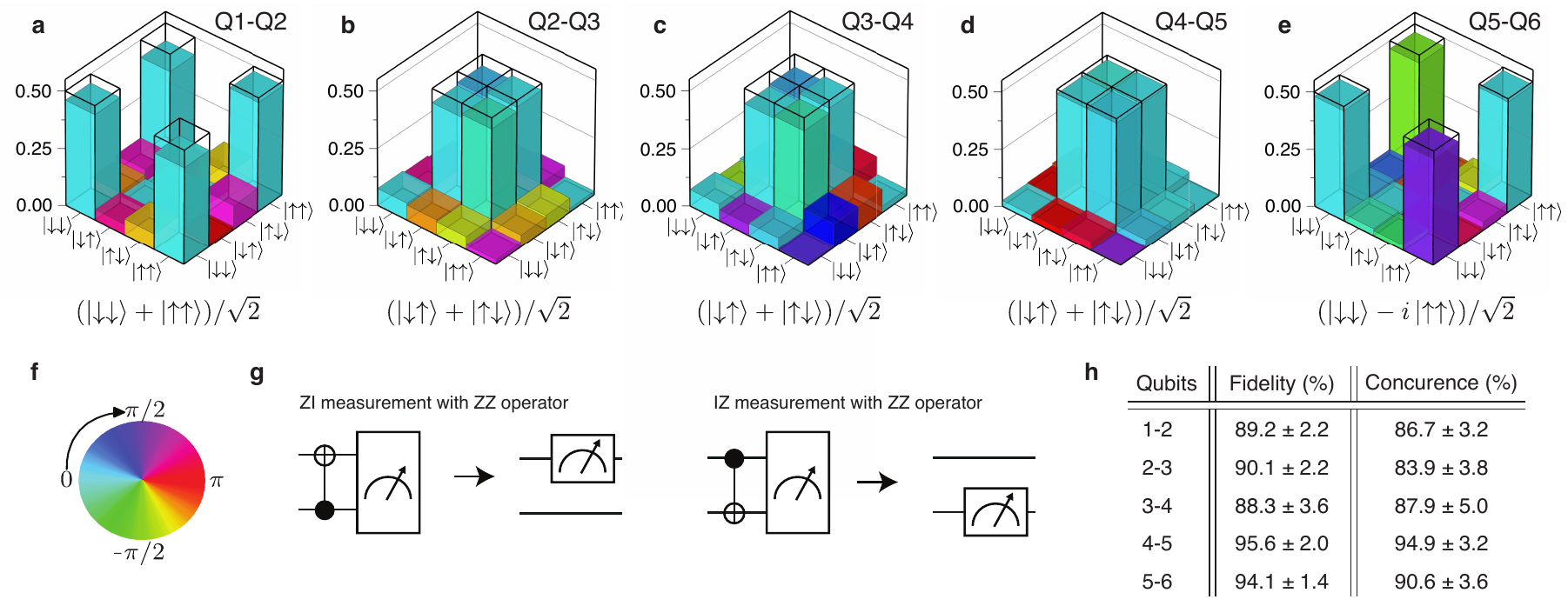}
	\caption{\textbf{Bell state tomography} \textbf{a-e}, Measured two-qubit density matrices for each pair of neighboring qubits, after removal of state preparation and measurement (SPAM) errors (see Supplementary Data for the uncorrected density matrices). The target Bell states are indicated and outlined with the wireframes. \textbf{f}, Colorwheel with phase information for the density matrices presented in panels \textbf{a-e}. \textbf{g}, Quantum circuits used for converting parity readout (ZZ) into effective single-qubit readout (IZ and ZI). \textbf{h}, State fidelities of the measured density matrices with respect to the target Bell states and the concurrences for the measured density matrices. Error bars ($2\sigma$) are derived from Monte-Carlo bootstrap resampling~\cite{simmons2011entanglement,huang2019fidelity,takeda2021quantum}. State fidelities without readout error removal, Q12: 88.2\%, Q23: 83.8\%, Q34: 78.0\%, Q45: 91.3\%, Q56: 91.3\%.} 
	\label{main_fig_3}
\end{figure*}

Two-qubit gates are implemented by pulsing the (virtual) barrier gate between adjacent dots while staying at the symmetry point. Pulsing the barrier gate leads to a ZZ interaction (throughout, X, Y and Z stand for the Pauli operators, I for the identity, and ZZ is shorthand for the tensor product of two Pauli Z operators, etc), given that the effect of the flip-flop terms of the spin exchange interaction is suppressed due to the differences in the qubit splittings~\cite{meunier2007experimental}. The quantum circuit in Fig.~\ref{fig:main_fig_2}a measures the time evolution under the ZZ component of the Hamiltonian only, as the single-qubit $\pi$ pulses in between the two exchange pulses decouple any IZ/ZI terms~\cite{watson2018programmable}. The measured signal oscillates at a frequency $J/2$ (Fig.~\ref{fig:main_fig_2}b-f) as a function of the barrier gate pulse duration, corresponding to Controlled Phase (CPhase) evolution. When pulsing only the barrier gate between the target qubit pair, the desired on/off ratio of $J_{\mathrm{ij}}$ (>100) could not be achieved. We solve this, without sacrificing operation at the symmetry point, by using a linear combination of the virtual barrier gates (vB1-vB6). Specifically, the barrier gates around the targeted quantum dot pair are pulsed negatively to push the corresponding electrons closer together and thereby enhance the exchange interaction (see Extended Data Fig. \ref{ED:Jpulses}). The exponential dependence of $J_{\mathrm{ij}}$ on virtual barrier gates is seen in Fig. \ref{fig:main_fig_2}h. In Fig.~\ref{fig:main_fig_2}g we investigate the residual exchange of idle qubit pairs, while one qubit pair is pulsed to its maximal exchange value within the operating range. The results show minimal residual exchange amplitudes in the off-state between the other pairs. 

Through suitable timing, we use the CPhase evolution to implement a Controlled-Z (CZ) gate. Fig.~\ref{fig:main_fig_2}j shows the pulse shape that is used to ensure a high degree of adiabaticity throughout the CZ gate~\cite{xue2021computing}. We use a Tukey window as waveform, with a ramp time of $\tau_{\mathrm{ramp}} = \frac{3}{\sqrt{\delta B^2 + J_{\mathrm{max}}^2}}$ \cite{martinis2014fast}. This pulse shape is defined in units of energy and we convert it into barrier voltages using the measured voltage to exchange energy relation~\cite{xue2021computing}.

One of the challenges when operating larger quantum processors is to track and compensate any dynamical changes in qubit parameters to ensure high-fidelity operation, initialization and readout. Another challenge is to keep track of and compensate for cross-talk effects imparted by both single- and two-qubit gates on the phase evolution of each qubit. We perform automated calibrations, as shown in Fig.~\ref{fig:main_fig_0}f, and correct 108 parameters in total. The detailed description of each calibration routine is included in the methods section and Extended Data Fig.~\ref{ED:calib}.  Twice a week, we run the full calibration scheme, which takes about one hour. Every morning, we run the calibration scheme leaving out the phase corrections for single-qubit operations and the dependence of $J_{\mathrm{ij}}$ on the virtual barrier gates vB$_{\mathrm{ij}}$. Sometimes, specific calibrations, especially qubit frequencies and readout coordinates, are rerun throughout the day, as needed. Supplementary Data figure 3 plots the evolution of the calibrated values for a number of qubit parameters over the course of one month.

With single- and two-qubit control established across the six-qubit array, we proceed to create and quantify pairwise entanglement across the quantum dot array as a measure of the quality of the qubit control (Fig.~\ref{main_fig_3}a-e). These experiments benefited from a high level of abstraction in the measurement software, allowing us to flexibly program a variety of quantum circuits acting on any of the qubits, drawing on the table of 108 calibration parameters that is kept updated in the background and on the detailed waveforms to achieve high-fidelity gates. The parity readout of the outer qubits yields a native ZZ measurement operator. We measure single-qubit expectation values by mapping the ZZ operator to a ZI/IZ operator, as shown in figure \ref{main_fig_3}g. This allows full reconstruction of the density matrix. The state fidelity is calculated using $F = \bra{\psi}\rho\ket{\psi}$, where $\psi$ is the target state and $\rho$ is the measured density matrix. The target states are maximally entangled Bell states. The obtained density matrices measured across the six-dot array have a state fidelity ranging from 88\% to 96\%, which is considerably higher than the Bell state fidelities of 78\% to 89\% reported on two-qubit quantum dot devices just a few years ago \cite{watson2018programmable, huang2019fidelity, zajac2018resonantly}. 

\begin{figure*}[!h]
	\includegraphics{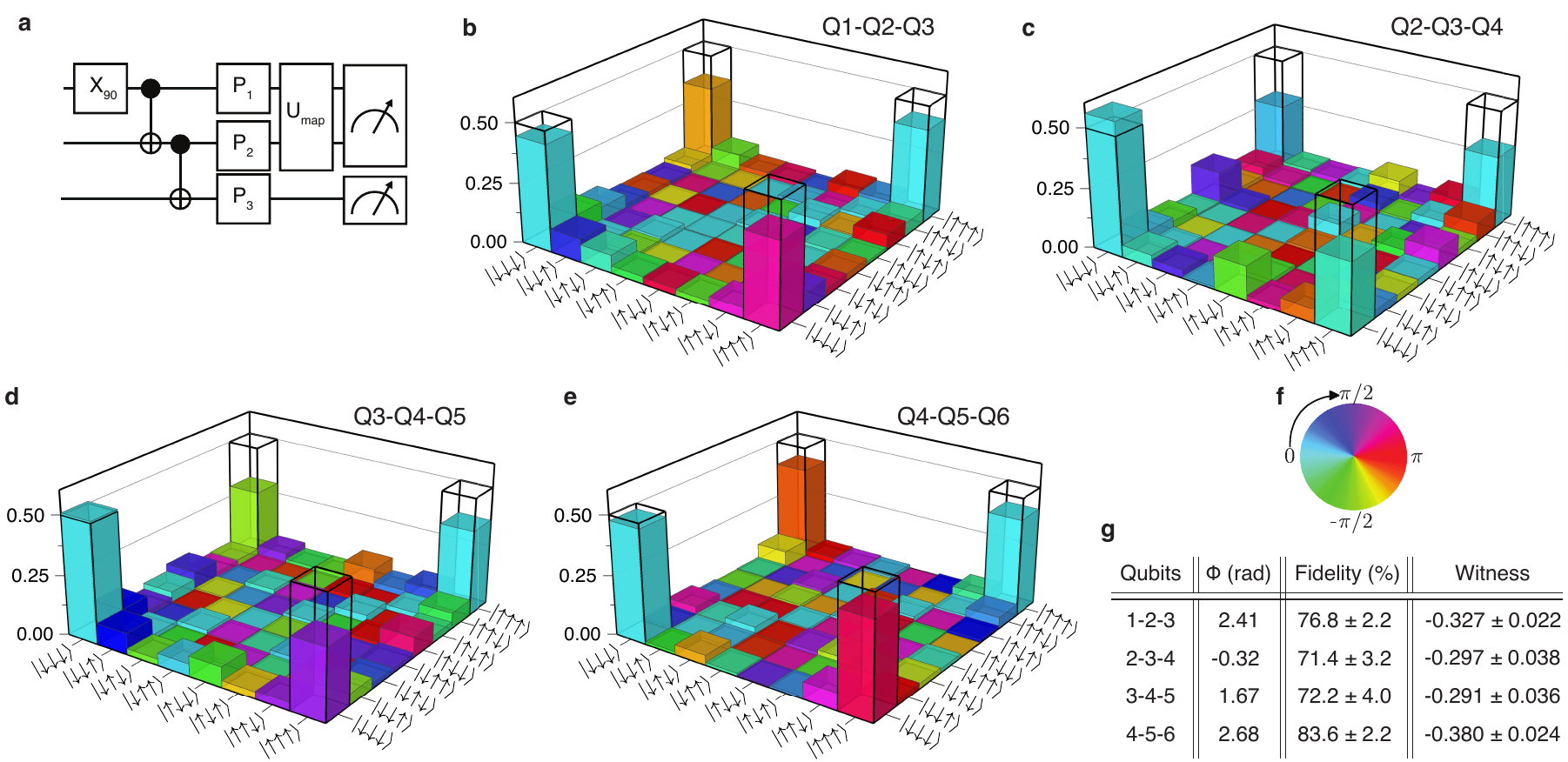}
	\caption{\textbf{Three-qubit Greenberger–Horne–Zeilinger state tomography} \textbf{a}, Circuit diagram used to prepare the GHZ states. The U$_{\mathrm{map}}$ operation is the unitary that is executed in case we measure the IZ or ZI projections on qubits 1/2 and 5/6, similar to the Bell state experiments. \textbf{b-e}, Density matrices of the prepared GHZ states using qubit 123, 234, 345 and 456, obtained using quantum state tomography, after removal of SPAM errors (see Supplementary Data for the uncorrected density matrices). The black wireframes correspond to the ideal GHZ state. \textbf{f}, Colorwheel with phase information for the density matrices presented in panels \textbf{d-e}. \textbf{g}, Table showing the state fidelities and entanglement witness values for the different qubit sets. We choose $\phi$ in $\ket{\psi_{\mathrm{GHZ}}}=(\ket{000} + e^{i\phi}\ket{111})\sqrt{2}$, with respect to the highest state fidelity.  State fidelities without SPAM removal, Q123: 64.3\%, Q234: 52.8\%, Q345: 52.7\%, Q456: 67.2\%}
	\label{main_fig_4}
\end{figure*}

As a final characterization of the qubit control across the array, we prepare  Greenberger–Horne–Zeilinger (GHZ) states, which are the most delicate entangled states of three qubits~\cite{greenberger1989going,rajagopal2002robust}. Fig.~\ref{main_fig_4}a shows the quantum circuit we used to prepare the GHZ states. The full circuit, including initialization and measurement, contains up to 14 CROT operations, 2 CZ operations, 42 parity measurements, 16 single-qubit rotations conditional on real-time feedback, and 5 single-qubit X$_{90}$ rotations (see Extended Data Fig.~\ref{ED:init_seq}). The measurement operators for quantum state tomography are generated in a similar manner as for the Bell states. In order to reconstruct three-qubit density matrices, we perform measurements in 26 (for qubits 234 and 345) or 44 (for qubit 123 and 456) different basis and repeat each set 2000 times to collect statistics. A full dataset consisting of 52000 (88000) single-shot repetitions takes about 5 minutes to acquire, thanks to the efficient uploading of waveforms to the waveform generator (see Methods) and the short single-shot cycle times. Figs.~\ref{main_fig_4}b-e show the measured density matrices for qubits 123, 234, 345 and 456. The obtained state fidelities range from 71\% to 84\% (see Methods for a brief discussion of dephasing effects from heating). For comparison, the record GHZ state fidelity reported recently for a triple quantum dot spin qubit system is 88\%~\cite{takeda2021quantum}. The same data set from~\cite{takeda2021quantum} analyzed without readout correction yields 45.8\% fidelity, while our results with no readout error removal range from 52.8\% to 67.2\% (see Supplementary Data). From the same data sets, we calculate entanglement witnesses, which clearly demonstrate three-qubit entanglement (see Supplementary Data). 

The demonstration of universal control of six qubits in a $^{28}$Si/SiGe quantum dot array advances the field in multiple ways. While scaling to a record number of qubits for a quantum dot system, we achieve Rabi oscillations for each qubit with visibilities of 93.5-98\%, implying high readout and initialization fidelities.
Initialization uses a novel scheme relying on qubit measurement and real-time feedback. Readout relies on Pauli spin blockade and quantum-non-demolition measurements. This combination of initialization and readout allows to the device to be operated while retaining the six electrons in the linear quantum dot array, alleviating the need for access to electron reservoirs. All single-qubit gate fidelities are around 99.9\% and the high quality of the two-qubit gates can be inferred from the 89-95\% fidelity Bell states prepared across the array. The development of a modular software stack, efficient calibration routines and reliable device fabrication have been essential for this experiment. Future work must focus on understanding and mitigating heating effects leading to frequency shifts and reduced dephasing times, as we find this to be the limiting factor in executing complicated quantum circuits on many qubits. The use of simultaneous single-qubit rotations and simultaneous two-qubit CZ gates will keep pulse sequences more compact. This will require accounting for cross-talk effects, which we anticipate will be easiest for the two-qubit gates. We estimate that the concepts used here for control, initialization and readout can be used without substantial modification in arrays that are twice as long, as well as in small two-dimensional arrays. Scaling further will require additional elements such as cross-bar addressing or on-chip quantum links~\cite{vandersypennpjQInfo}.  


\section*{Methods}
\subsection*{Device fabrication}
Devices are fabricated on an undoped $^{28}$Si/SiGe heterostructure featuring an 8~nm strained $^{28}$Si quantum well, with a residual $^{29}$Si concentration of 0.08\%, grown on a strain-relaxed Si$_{0.7}$Ge$_{0.3}$ buffer layer. The quantum well is separated from the surface by a 30~nm thick Si$_{0.7}$Ge$_{0.3}$ spacer and a sacrificial 1~nm Si capping layer. 
The gate stack consists of 3 layers of Ti:Pd metallic gates  (3:17, 3:27, 3:27~nm) isolated from each other by 5~nm Al$_2$O$_3$ dielectrics, deposited using atomic layer deposition. 
A ferromagnetic Ti:Co (5:200~nm) layer on top of the gate stack creates a local magnetic field gradient for qubit addressing and manipulation.
Further details of device fabrication methods can be found at \cite{lawrie_quantum_2020}.

\subsection*{Microwave crosstalk and synchronization condition}
In Fig.~\ref{fig:main_fig_1}, the single-qubit gates are chosen to be operated at a 5 MHz Rabi frequency and all single-qubit randomized benchmarking (RB) results are taken at this frequency as well. When operating all qubits within the same sequence, we were unable to operate at a 5 MHz Rabi frequency as qubits 2(3) and 5(4) are too close to each other in frequency. We used the synchronization condition~\cite{heinz2021crosstalk,russ2018high} to choose Rabi frequencies for the single-qubit gates for which the qubit that suffers cross-talk does not undergo a net rotation while the target qubit is rotated by 90 degrees or multiples thereof (see Extended Data figure \ref{ED:crosstalk}). The Rabi frequencies for the state tomography experiments are as follows (qubit 1-6): 4.6~MHz, 1.9~MHz, 4.2~MHz, 3.6~MHz, 2.4~MHz and 5~MHz.

\subsection*{Automated calibration routines}
Calibrations are a crucial part in operating a multi-qubit device. Figure \ref{fig:main_fig_0}d list the necessary parameters that need to be corrected periodically and the Extended Data Fig. \ref{ED:calib} shows an example calibration for each parameter type. In each panel, the value extracted in the corresponding calibration is indicated.

We perform the full calibration routine twice a week at most, and throughout the day we execute parts of the calibration protocol, when we suspect a parameter has drifted (e.g. when we observe a reduced visibility). For each calibrated parameter, an automated script detects the optimal value and updates the record in the variable manager. In our framework, the operator chooses to accept this value or to re-run the calibration.

\paragraph{Sensing dot -- (5~s) --}
The calibrations routine starts by calibrating the sensing dots (Extended Data Fig. \ref{ED:calib}a) to the most sensitive operating point for parity mode PSB readout. We scan the (virtual) plunger voltage of the sensing dot for two different charge configurations of the corresponding double dot, corresponding to the singlet and triplet states. One configuration is well in the (3,1) region, the other well in the (4,0) region, in order to be insensitive to small drifts in the gate voltages. The calibration returns the plunger voltage for which the largest difference is obtained in the sensing dot signal between these two cases (Extended Data Fig. \ref{ED:calib}a). From this difference, we also set the threshold in the demodulated IQ signal of the RF-readout, to allow singlet/triplet differentiation (the IQ signal is converted to a scalar by adjusting the phase of the signal). The threshold is chosen halfway between the signals for the two charge configuration. During qubit manipulation, the sensing dot is kept in Coulomb blockade. It is only pulsed to the readout configuration when executing the readout.

\paragraph{Readout point -- (35~s) --}
The parity mode PSB readout is calibrated by finding the optimal voltage of the plunger gates near the anticrossing for the readout. The readout point is only calibrated along one axis (vP1 or vP5), for simplicity, and since the performance of the PSB readout is similar at any location along the anticrossing. In the calibration shown in Extended Data Fig. \ref{ED:calib}b, we initialize either a singlet ($\ket{\uparrow\downarrow}$) or a triplet ($\ket{\downarrow\downarrow}$, using a single-qubit gate) state and sweep the plunger gate to find to optimal readout point. 

\paragraph{Q1/2/5/6 resonance frequency (rough)  -- (17~s) --}
We perform a course scan of the resonance frequencies of qubits 1, 2, 5 and 6 (Extended Data Fig. \ref{ED:calib}c) around the previously saved values. We fit the Rabi formula 
\begin{align}
    P_s(t) = \frac{\Omega^2}{\Omega^2 + \Delta^2}\sin^2 \left( \frac{\sqrt{\Omega^2 + \Delta^2}}{2}t \right)
\end{align}
to the experimental data and extract the resonance frequency.

\subsubsection*{QND readout: CROT Q32/Q45 $f_{res}$ -- (14~s) --}
Subsequently, we calibrate the QND readout for qubits 3 and 4. To perform QND readout, we need to calibrate a CROT gate. We choose to use a controlled rotation two-qubit gate, as it requires little calibration (compared to the CPhase) given that we can ignore phase errors during readout.

We set the exchange to 10-20 MHz via barrier gate pulses and scan the CROT driving frequency (Extended Data Fig. \ref{ED:calib}d) around the previously saved values. Again, we fit the Rabi formula 
\begin{align}
    P_s(t) = \frac{\Omega^2}{\Omega^2 + \Delta^2}\sin^2 \left( \frac{\sqrt{\Omega^2 + \Delta^2}}{2}t \right)
\end{align}
to extract the optimal resonance frequency.

\paragraph{QND readout: CROT Q32/Q45 pulse width -- (25~s) --}
Next, we tune the optimal microwave burst duration for the CROT gate, by driving Rabi oscillations (Extended Data Fig. \ref{ED:calib}e) in the presence of the exchange coupling. We fit the decaying sinusoid 
\begin{align}
    P_s(t) = \frac{A}{2}\sin(\omega t - \phi_0)e^{-\frac{t}{\tau}} + B
\end{align}
and extract the pulse width the for CROT gate.

\paragraph{Q3/4 resonance frequency (rough)  -- (28~s) --}
With QND readout established, we scan the driving frequency for qubit 3 and 4 in a similar manner as we did for Q1/2/5/6 (Extended Data Fig \ref{ED:calib}f). The calibration scripts will automatically use QND readout for Q3 and Q4 calibration, in place of PSB readout for Q1, Q2, Q5, Q6.

\paragraph{Q1-6 resonance frequency and amplitude (fine) -- (Q1, Q2, Q5, Q6 $\rightarrow$ frequency 22~s, amplitude 23~s; Q3, Q4 $\rightarrow$ frequency 32~s, amplitude 34~s) --}
We calibrate more accurately the qubit frequency and driving amplitude using an error amplification sequence (Extended Data Fig. \ref{ED:calib}g-h), where we execute an X$_{90}$ gate 18 times and sweep either the frequency or the amplitude of the microwave burst. We fit the data using the Rabi formula once again
\begin{align}
    P_s(t) = \frac{\Omega^2}{\Omega^2 + \Delta^2}\sin^2 \left( \frac{\sqrt{\Omega^2 + \Delta^2}}{2}t \right)
\end{align}
to extract the resonance frequency. The amplitude of the microwave burst is controlled via the I/Q input channels of the vector source we used. To calibrate the amplitude for an X$_{90}$ rotation, we vary the amplitude applied to the I/Q input and fit the result to
\begin{align}
    P_s(x) = \alpha e^{-\frac{(x-\mu)^2}{2\sigma^2}} \;.
\end{align}
This functional form is not strictly correct but it does find the optimal amplitude for an X$_{90}$ rotation. We suspect that the longer amplification sequences gave better results, as they more closely resemble the sequence lengths used for randomized benchmarking (including some 'heating effects').

In these calibrations, we only calibrate the X$_{90}$ gate. The Y$_{90}$ gate is implemented similarly to the X$_{90}$, but phase shifted. Z gates are performed in software by shifting the reference frame. X$_{180}$ and Y$_{180}$ rotations are performed by applying two 90 degree rotations. We do not simultaneously drive two or more qubits. 

\paragraph{Q1-6 X$_{90}$ phase crosstalk -- (Q1, Q2, Q5, Q6 $\rightarrow$ 27~s; Q3, Q4 $\rightarrow$ 45~s) --}
Any single-qubit gate causes the Larmor frequency of the other qubits to shift slightly due to  the applied microwave drive. We compensate for this by applying a virtual Z rotation to every qubit after a single-qubit gate has been performed. The Ramsey based sequence is used to calibrate the required phase corrections (Extended Data Fig.~\ref{ED:calib}i) and data is fitted with
\begin{align}
    P_s(\phi) = -\frac{A}{2}\cos(\phi - \phi_0) + B
\end{align}
to extract the necessary phase correction, $\phi_0$. A single X$_{90}$ pulse on one qubit will impart phase errors on qubits 2 to 6. Thus we need to calibrate separately 30 different phase factors, five for each qubit.

\paragraph{\textit{J}$_{\mathrm{ij}}$ vs vB$_{\mathrm{ij}}$ -- (Qubit pairs 12, 56 $\rightarrow$ 146~s; Qubit pairs 23, 45 $\rightarrow$ 207~s; Qubit pair 34 $\rightarrow$ 299~s) --}
Two-qubit gates are implemented by applying a voltage pulse that increases the tunnel coupling between the respective quantum dots. To enable two-qubit gates, we take the following elements into account:
\begin{itemize}
    \item Exchange strength. We operate the two-qubit gates at exchange strengths $J_{\mathrm{on}}$ where the quality factor of the oscillations is maximal. This condition is found for $J_{\mathrm{on}} \approx 5$ MHz.
    \item Adiabacity condition. When the Zeeman energy difference ($\Delta E_{\mathrm{z}}$) and the exchange ($J(t)$) are of the same order of magnitude, care has to be taken to maintain adiabaticity throughout the CPhase gate. We do this by applying a Tukey based pulse, where the ramp time is chosen as $\tau_{\mathrm{ramp}} = \frac{3}{\sqrt{\Delta E_{\mathrm{z}}^2 + J_{\mathrm{on}}^2}}$~\cite{martinis2014fast}.
    \item Single-qubit phase shifts. As we apply the exchange pulse, the qubits will physically be slightly displaced. This causes a frequency shift and hence phase accumulation, which needs to be corrected for.
\end{itemize}
In order to satisfy these conditions, we need to know the relationship between the barrier voltage and the exchange strength. We construct this relation by measuring the exchange strength (see figure \ref{fig:main_fig_2}a) for the last 25\% of the virtual barrier pulsing range ($J>1$~MHz regime). We fit the exchange to an exponential and extrapolate this to any exchange value (Extended Data Fig. \ref{ED:calib}j). This allows us to generate the adiabatic pulse as described in the main text and choose the target exchange value.

\paragraph{CZ duration -- (Qubit pairs 12, 56 $\rightarrow$ 29~s; Qubit pairs 23, 45 $\rightarrow$ 34~s; Qubit pair 34 $\rightarrow$ 45~s) --}
The gate voltage pulse to implement a CZ operation uses a Tukey shape in $J$ by inverting the relationship $J$(vB$_{\mathrm{ij}}$). The maximum value of $J$ is capped at $J_{\mathrm{on}}$. The actual largest value of $J$ used and the length of the pulse then determine the phase acquired under ZZ evolution. We first analytically evaluate the accumulated ZZ evolution as a function of these parameters around the target of $\pi$ evolution under ZZ, and then experimentally fine tune the actual accumulated ZZ evolution by executing a Ramsey circuit with a decoupled CPhase evolution in between the two $\pi/2$ rotations. An example of such a calibration measurement is shown in Extended Data Fig.~\ref{ED:calib}k.

\paragraph{CZ phase crosstalk -- (Q1, Q2, Q5, Q6 $\rightarrow$ $<$30~s; Q3, Q4 $\rightarrow$ $<$50~s) --}
After the exchange pulse is executed, single-qubit phases have to be corrected. We correct these phases on all the qubits, whether participating or not in the two-qubit gate. We calibrate the required phase corrections in a very similar way as done for the single-qubit gate phase corrections. An example of the circuit and measurement is given in Extended Data Fig.~\ref{ED:calib}l-m. The exact calibration run-time depends on the CZ pulse width and can vary by couple of seconds depending on the target qubit.

\subsection*{Heating effects}
We observed several effects that bear a signature of heating in our experiments. When microwaves (MWs) are applied to the EDSR line of the sample, several qubit properties change by an amount that depends on the applied driving power and the duty cycle of applying power versus no power. This effect has also been observed in other works \cite{takeda2018optimized}. We report our findings in Extended Data Fig.~\ref{ED:heating} and will discuss adjustments made to the sequences of the experiments to reduce their effects. The main heating effects are a reduction of the signal-to-noise ratio (SNR) of the sensing dot and a change of the qubit resonance frequency and T$_2^*$.

In Extended Data Fig.~\ref{ED:heating}a-d, we investigate the effect of a MW burst applied to the EDSR driving gate, after which the signal of the sensing dot is measured. We observe changes in the background signal and in the peak signal (the electrochemical potential of the sensing dot is not affected, as the peak does not shift in gate voltage). Since the background signal rises more than the peak signal, the net signal is reduced. This reduction depends on the magnitude and duration of the applied MW pulse (Extended Data Fig.~\ref{ED:heating}b). The original SNR can be recovered by introducing a waiting time after the MW pulse. The typical timescale needed to restore the SNR is in the order of 100~\textmu s (see Extended Data Fig.~\ref{ED:heating}c-d). We added for all (RB) data taken in this paper a waiting of 100~\textmu s (500~\textmu s) after the manipulation stage to achieve a good balance between SNR and experiment duration. Spin relaxation between manipulation and readout is negligible, given that no T$_1$ decay was observed on a timescale of 1~ms within the measurement accuracy. We did not introduce extra waiting times after feedback/CROT pulses in the initialization/readout cycle, as the power to perform these pulses did not limit the SNR.

Extended Data Fig.~\ref{ED:heating}f gives more insight in what makes the background and peak signal of the sensing dots change. The impedance of the sensing dot is measured using RF-reflectometry. The background of the measured signal depends on the inductance of the surface-mount inductor, the capacitance to ground\cite{connors2020rapid, noiri2020radio, liu2021radio} and the resistance to ground of the RF readout circuit.  Extended Data Fig.~\ref{ED:heating}f shows the response of the readout circuit under different MW powers (the RF power is kept fixed). A frequency shift (0.5~MHz) and a reduction in quality factor is observed. This can be indicative of an increase in capacitance and dissipation in the readout circuit. Presently the microscopic mechanisms that cause this behavior are unknown.

The second effect is observed when looking at the qubit properties themselves. Extended Data Fig.~\ref{ED:heating}e shows that both the dephasing time T$_2^*$ measured in a Ramsey experiment and the qubit frequency are altered by the microwave radiation. In the actual experiments, we apply a MW pre-pulse of 1-4~\textmu s before the manipulation stage to make the qubit frequency more predictable, though this comes at the cost of a reduced T$_2^*$. The pre-pulse can be applied either at the start or at the end of the pulse sequence, with similar effects. This indicates that heating effects on the qubit frequency persist for longer than the total time of a single-shot experiment ($\approx$ 600~\textmu s), different from the effect on the sensing dot signal. Also the microscopic mechanisms behind the qubit frequency shift and T$_2^*$ reduction remain to be understood.

\subsection*{Parity mode PSB readout}
Pauli spin blockade (PSB) readout is a method used to convert a spin state to a more easily detectable charge state\cite{ono2002current}. Several factors need to be taken into account for this conversion, to enable good readout visibilities. Extended Data Fig.~\ref{ED:parity}a-b shows the energy level diagrams for PSB readout performed in the (1,1) and (3,1) charge occupation. The diagrams use valley energies of 65~\textmu eV, to illustrate where problems can occur. When looking at Extended Data Fig.~\ref{ED:parity}, we can observe two potential issues:
\begin{itemize}
    \item The excited valley state with $\ket{\downarrow \downarrow}$ is located below the ground valley state with $\ket{\uparrow \downarrow}$. We assume in the diagram that the (2,0) singlet state ($\ket{\mathrm{S,0}}$) is coupled to both the (1,1) ground valley state and the (1,1) excited valley state. In this case, during the initialisation/readout pulses, population can be moved into the excited valley state. This problem can be solved by working at a lower magnetic field, such that $E_{\mathrm{v}} > E_{\mathrm{z}}$ (panel Extended Data Fig.~\ref{ED:parity}b).
    \item When operating in the (1,1) charge occupation, the readout window is quite small, as the size is determined by the difference between the valley energy and the Zeeman energy. A common way to prevent this problem is by operating in the (3,1) electron occupation.
\end{itemize}
With both measures in place, we consistently obtain high visibilities of Rabi oscillations ($\geq$ 94\%) on every device tested.

In the following we describe the procedure used to tune up the parity mode Pauli Spin Blockade.
\begin{enumerate}
    \item Find an appropriate tunneling rate at the (3,1) anticrossing. An initial guess of a good tunneling rate can be found using video mode tuning. We use the arbitrary waveform generator to record at high speed frames of the charge stability diagram (5~\textmu s averaging per frame, $t_{\mathrm{image}} = 200$~ms). While the frames are measured, we vary the tunnel coupling, while looking at the  (3,1) $\leftrightarrow$ (4,0) anticrossing until the pattern shown in Extended Data Fig.~\ref{ED:parity}c is observed. This figure shows that depending on the (random) initial state, the transition from (3,1) to (4,0) occurs at either location (i) or location (ii). This is exactly what needs needs to happen when the readout is performed. 
    \item Find the readout point. We hold point (1) fixed in the center of the (3,1) charge occupation (Extended Data Fig.~\ref{ED:parity}c). Point (2) is scanned with the AWG along the detuning axis as shown in Extended Data Fig.~\ref{ED:parity}c. We pulse from point (1) to point (2) and measure the state (ramp time of $\sim$2~\textmu s), then we pulse back to point (1). When plotting the measured singlet probability, a gap is seen between the case where a singlet is prepared versus a random spin state is prepared (Extended Data Fig.~\ref{ED:parity}d). The center of this region is a good readout point.
    \item Optimizing the readout parameters. The main optimization parameters are the detuning ($\epsilon$), tunnel coupling ($t_{\mathrm{c}}$) and ramp time to ramp towards the PSB region. We also independently calibrate the ramp time and tunnel coupling from the readout zone towards the operation point of the qubits. When ramping in towards the readout point, it is important to be adiabatic with respect to the tunnel coupling. We do not need to be adiabatic with respect to spin, as both $\ket{\uparrow\downarrow}$ and $\ket{\downarrow\uparrow}$ relax quickly to the singlet state (faster than we can measure, $<$1~ns). When pulsing from the readout to the operation point, more care has to be taken. When using the ramp time which performs well for readout, we notice that we initialize a mixed state, as we are not adiabatic with respect to spin. This can be solved by pulsing the tunnel coupling to a larger value before initiating the initialization ramp (Extended Data Fig~\ref{ED:parity}g).
\end{enumerate}
We show in the Extended Data Fig.~\ref{ED:parity}e-f, that the histograms for parallel and anti-parallel spin states are well separated, allowing for a spin readout fidelity exceeding 99.97\% for both qubits 1-2 and for qubits 5-6. This number could be further increased by integrating the signal for longer, but is not the limiting process. This way of quantifying the spin readout fidelity is commonly used in the literature but it leaves out errors occurring during the ramp time (mapping of qubit states to the readout basis states). This can be a significant effect, as seen from the measured visibility of the Rabi oscillations.

\subsection*{Setup and the Real-time feedback using FPGA}
\paragraph{Setup} 
A detailed schematic of the experimental setup is presented in Extended Data Fig.~\ref{ED:setup}, listing all the key components used in the experiment. 

\paragraph{Programming quantum circuits}

The quantum circuits are implemented in the form of microwave bursts for single-qubit operations, gate voltage pulses for two-qubit gates, and gate voltage pulses combined with RF bursts for readout. The gate voltage pulses are generated by an arbitrary wave generator (AWG). The microwave bursts are generated through IQ modulation of a MW vector source carrier frequency. The input signals for the IQ modulation are generated by the same AWG as used for the voltage pulses. The IQ modulation defines the amplitude envelope of the microwave bursts, the output frequency and the phase shifts. Virtual-Z gates are implemented by incrementing the reference phase of the NCO (see below) and are used to e.g. correct phase errors introduced by crosstalk. The generated control signals are stored in memory with a resolution of 1~ns. 

Microwave bursts applied to the six qubit sample are supplied by a single MW source with a carrier frequency set at 16.3~GHz. We address the six different qubits using single side-band IQ modulation of the carrier to displace the frequency of the microwave output signal to the frequency of the target qubit. As each qubit has a different resonance frequency (and different from carrier frequency), it is necessary to track the phase evolution at the qubit Larmor precession frequency to ensure phase coherent MW bursts for successive single-qubit operations. To realize that, we define in the AWG six continuously running numerically controlled oscillators (NCO), one for each qubit. These NCOs keep track of the qubits' phase evolution with respect to the carrier frequency. We choose this approach instead of pre-calculating phase factors for every pulse in a sequence, which is a not a scalable approach with a growing complexity of the quantum circuits. 

The digitizer is synchronised with the AWG to acquire qubit readout data. In a single-shot we can include multiple readout segments, each defined in a digitizer instruction list. A step in this list specifies a measurement time window, a wait time and the threshold for the qubit state. The input signal is integrated during the measurement window and the result is compared with a threshold to determine the qubit state. This outcome, 0 or 1, can be passed directly to the AWG via a PXI trigger line shared by digitizer and AWGs, to realize real-time feedback on the measurement output.

\paragraph{Real-time feedback}
In the initialization and readout sequences the execution of selected gates depends on the outcomes of intermediate measurements, allowing for real-time qubit state corrections. The total time from the end of the measurement until the start of the conditional gate (burst) on the device should be much shorter than the qubit relaxation time T$_1$, and ideally also shorter than $\sim$1~\textmu s -- the time needed for the adiabatic passage back to the manipulation point after the parity measurement -- such that no unnecessary idling time is spent. This fast control loop is realized with a custom FPGA program in the arbitrary wave generator (AWG) and digitizer as shown in Extended Data Fig.~\ref{ED:feedback}. The total latency for the closed loop feedback is 660~ns, which fits the design requirements.

\section*{Data availability}
The raw data and analysis that support the findings of this study are available in the Zenodo repository (https://doi.org/10.5281/zenodo.6138474)

\section*{Code availability}
The measurement and analysis code is available in the Zenodo repositories (core-tools https://zenodo.org/badge/latestdoi/264858832; pulse library https://zenodo.org/badge/latestdoi/113251242; qubit abstraction layer  https://zenodo.org/badge/latestdoi/253903530; state-tomography https://zenodo.org/record/6135943).

\bibliographystyle{naturemag}
\bibliography{bibliography.bib}

\section*{Acknowledgments}
We acknowledge R. Schouten for general advice and help on the measurement electronics, M. Almendros and his team for a collaborative development of FPGA hardware control, H. Van Der does and N. Philips for the design of the sample printed circuit board, Z. Jiang, A.-M. Zwerver, L. Peters and F. Unseld for assistance with the testing of samples, M. Eriksson and his team for contributions to sample fabrication, and members of the Vandersypen group for useful discussions. We acknowledge financial support from the Marie Skłodowska-Curie actions—Nanoscale solid-state spin systems in emerging quantum technologies—Spin-NANO, grant agreement number 676108. This research was sponsored by the Army Research Office (ARO) under grant numbers W911NF-17-1-0274 and W911NF-12-1-0607. The views and conclusions contained in this document are those of the authors and should not be interpreted as representing the official policies, either expressed or implied, of the ARO or the US Government. The US Government is authorized to reproduce and distribute reprints for government purposes notwithstanding any copyright notation herein. Development and maintenance of the growth facilities used for fabricating samples is supported by DOE (DE-FG02-03ER46028). We acknowledge support from Keysight’s University Research Collaborations.

\section*{Author contributions}
S.G.J.P. and M.T.M. performed the experiment with help from C.V. Data analysis was carried out by S.G.J.P., M.T.M. and M.R., who also performed the numerical simulations of the Bell and GHZ states.
S.G.J.P. and S.L.S. wrote the libraries used to control the experiment. S.L.S. wrote the library used for real-time feedback and made the supporting FPGA images. 
S.G.J.P., M.T.M., M.R. and L.M.K.V. contributed to the interpretation of the data. S.V.A., N.K., D.B., W.I.L.L., M.V. and L.T contributed to device fabrication and A.S., B.P.W and G.S designed and grew the Si/SiGe heterostructure. S.G.J.P., M.T.M. and L.M.K.V. wrote the manuscript with comments by all authors. L.M.K.V. conceived and supervised the project.

\newpage
\onecolumn
\section*{Extended Data figures and tables}
\renewcommand{\figurename}{Extended Data Fig.}
\setcounter{figure}{0}    

\begin{center}
\includegraphics{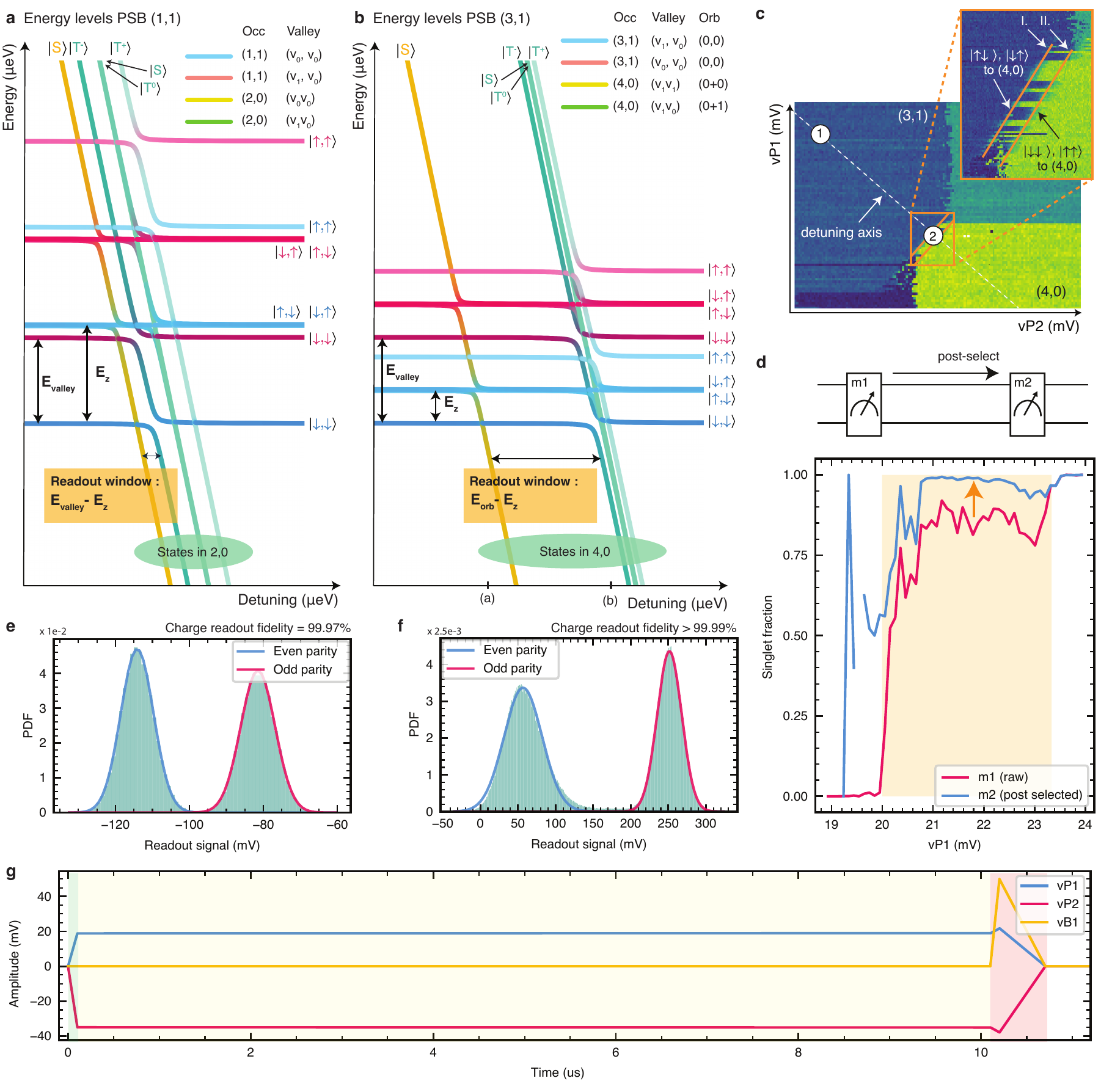}
    \captionof{figure}{\textbf{Pauli Spin Blockade readout} 
    \textbf{a}, Energy diagram for a double quantum dot as a function of the detuning between the (1,1)/(2,0) charge electron occupation. The Zeeman energy ($E_{\mathrm{z}}$) for qubit 1/2 is set to 74/75~\textmu eV (18/18.2~GHz) and the valley energy ($\epsilon_{\mathrm{v}}$) of dot 1 is set to 65~\textmu eV. We set $\epsilon_{\mathrm{c}}$ for the second dot to a much larger value to shift part of the energy spectrum out of view and simplify the visual analysis. The charge and valley occupations are indicated in the top right of this panel. \textbf{b}, Energy diagram in the (3,1)/(4,0) charge occupation. This panel uses identical parameter values as panel \textbf{a}, except for the Zeeman energy for qubit 1 and 2, 25/26~\textmu eV (6 and 6.2~GHz). The excited state energy of dot 1 in the (4,0) charge occupation is given by the orbital energy instead of the valley splitting. \textbf{c} Experimental charge stability diagram taken at the (3,1)/(4,0) anticrossing for dots 1 and 2. The point indicated with '1' indicates the qubit operation point and the point indicated as '2' indicates the readout point. The inset zooms in on the anticrossing, allowing one to observe the spin selective tunneling for the different input states (the readout zone). \textbf{d} In our experiment we initialize via measurement and post-selection. In the plot we can see the effects of two subsequent readouts. First readout 'm1 (raw)' shows the initial singlet fraction (electron spins are not intentionally randomized and by nature of the executed measurement sequence, a singlet state is preserved for next single shot). Second measurement 'm2' shows the outcome from post-selection on the result of 'm1' (realized as per-measurement-point post-processing in software). Within the readout zone (shaded area) the initialized singlet fraction is greatly amplified. \textbf{e-f} Probability density function of the PSB readout signal between qubit pairs 12 and 56 (10~\textmu s integration time), recorded in the course of the Rabi oscillations of qubit 1/5 in figure \ref{fig:main_fig_1}. The Gaussians are fitted to the two distributions and charge readout fidelities, estimated from their overlap, exceeding 99.9\%. No T$_1$ decay was observed at the readout point (T$_1>>$ 100~\textmu s). \textbf{g}, Gate voltage pulses applied to perform PSB readout on qubit pair 12. The different background colors indicate the ramp towards the readout point (green), measurement at the readout point (yellow) and the adiabatic ramp back to the operating point (red).
    }
	\label{ED:parity}
\end{center}

\begin{center}
\includegraphics{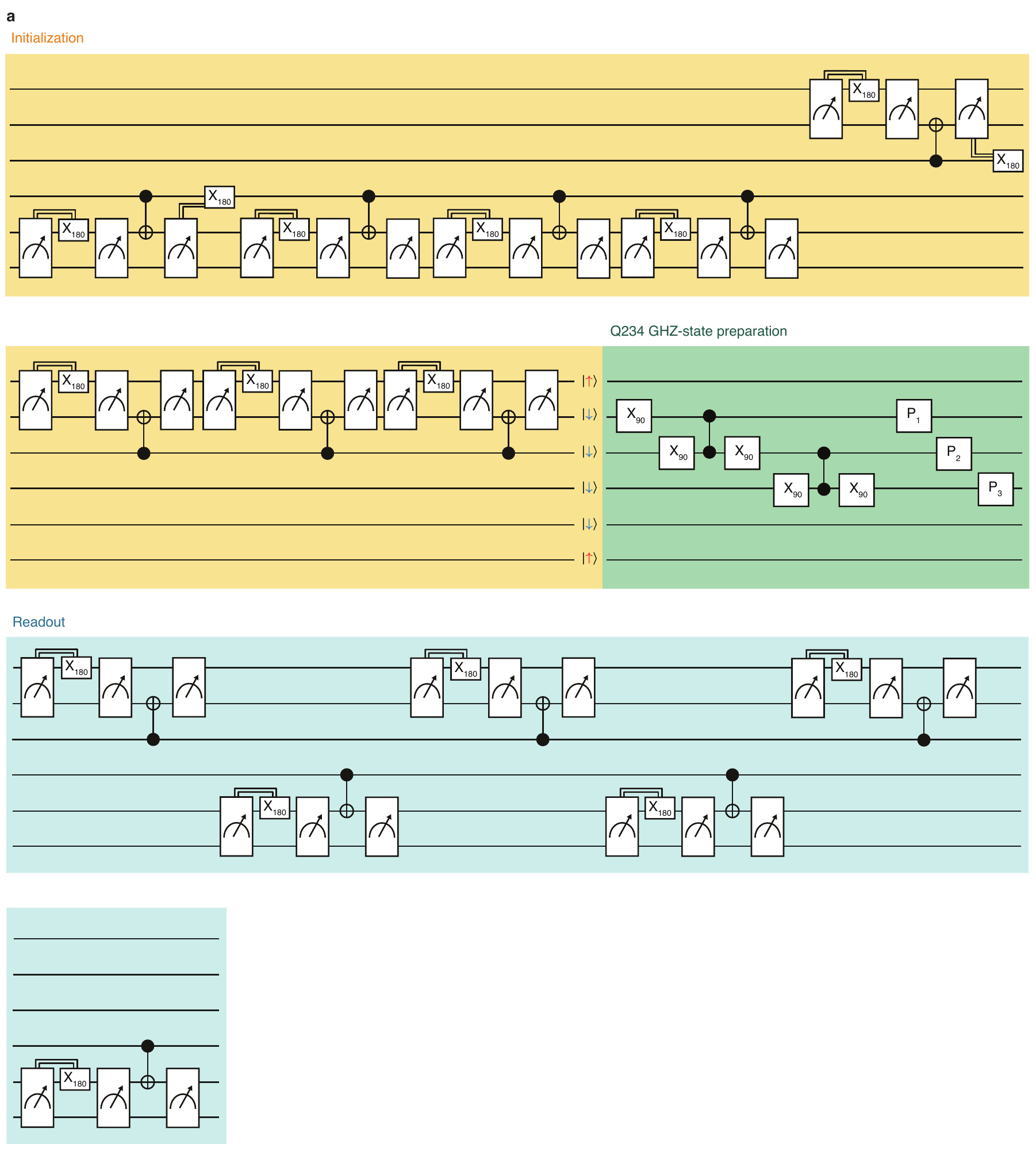}
    \captionof{figure}{\textbf{Full quantum circuits.}
    \textbf{a}, This circuit was used to produce the data presented in Fig.~\ref{main_fig_4}c. The different background colors indicate the different parts of the sequence (yellow -- initialization, green -- manipulation (including the tomography pulses), blue -- readout).}
	\label{ED:init_seq}
\end{center}

\begin{center}
\includegraphics{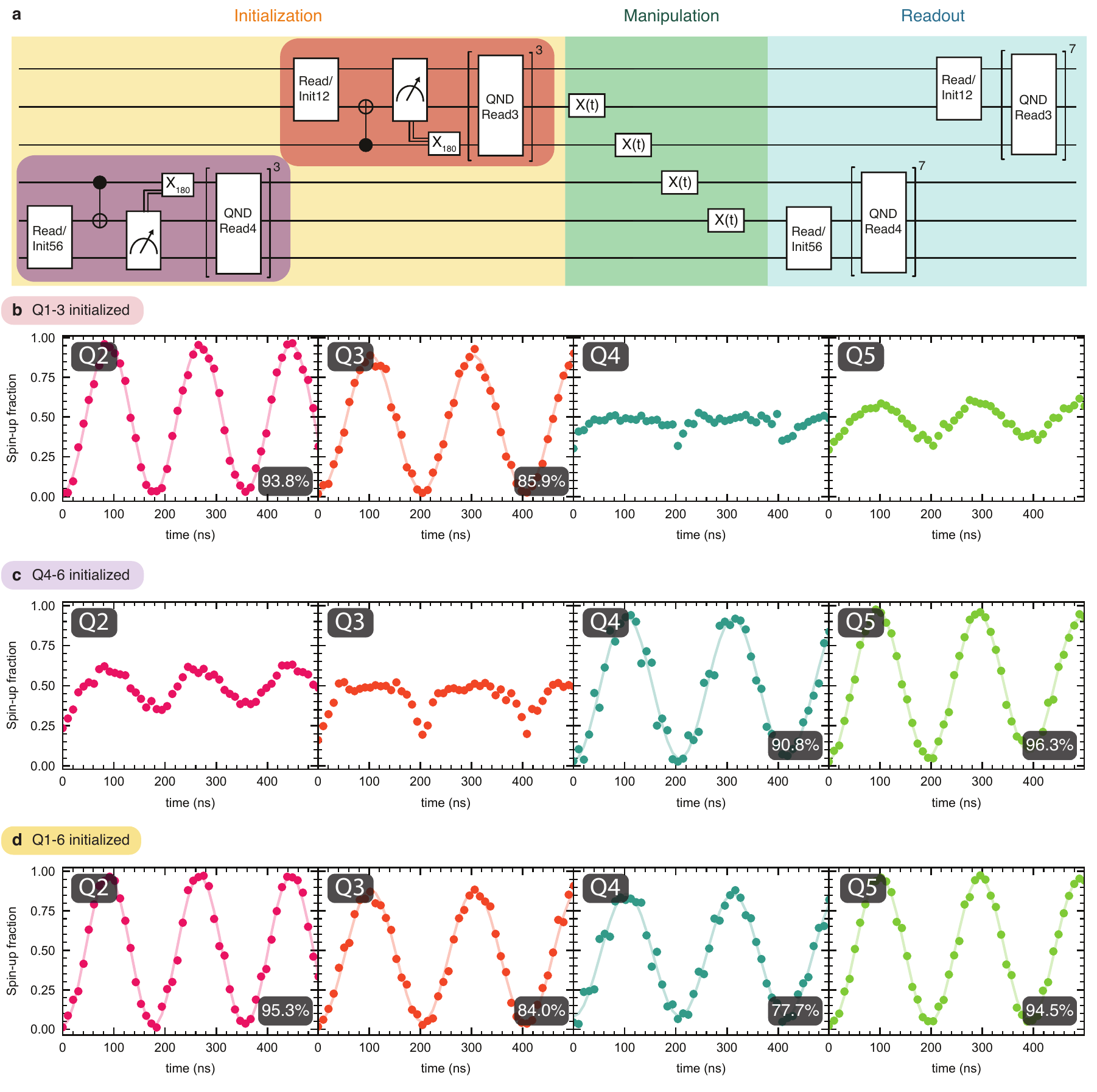}
    \captionof{figure}{\textbf{Loss in visibility when initializing all qubits.} 
    \textbf{a}, Gate sequence used to demonstrate the effect of running different qubit initialization routines.  Microwave bursts with variable duration are performed sequentially on qubits 2, 3, 4 and 5, as shown in the schematic. In all cases, readout is performed on all the qubits, but we vary which qubits are initialized for a particular experiment. \textbf{b-d}, Results of the sequence displayed in panel \textbf{a}, when qubit 1-3 (\textbf{b}),  4-6 (\textbf{c}) or 1-6 (\textbf{d}) are initialized. The shaded numbers in the panels indicate the visibility of the measured qubit. With all qubits initialized we observe a visibility loss on qubit 4 and to a lesser extent qubit 3. A visibility loss in principle can originate from a reduced readout or initialization fidelity. We keep the readout sequence identical for \textbf{a, b,} and \textbf{c}. Additionally, we include a 500~\textmu s waiting time prior to readout, to minimize any effects of MW pulsing during initialization or manipulation stage on the readout performance (see Methods). Although, we cannot be certain that the readout fidelity is unaffected by the initialization of all qubits, we speculate that the majority of the observed visibility loss is due to a reduced initialization fidelity, possibly due to the sensitivity of the CROTs to qubit frequency shifts. This interpretation is consistent with the fact that mostly qubit 4 suffers a lower visibility, as qubit 4 is initialized before qubits 1,2 and 3. If instead we reverse the initialization order, qubit 3 displays lowered visibility (data not shown).}
	\label{ED:visibility}
\end{center}

\begin{center}
    \includegraphics{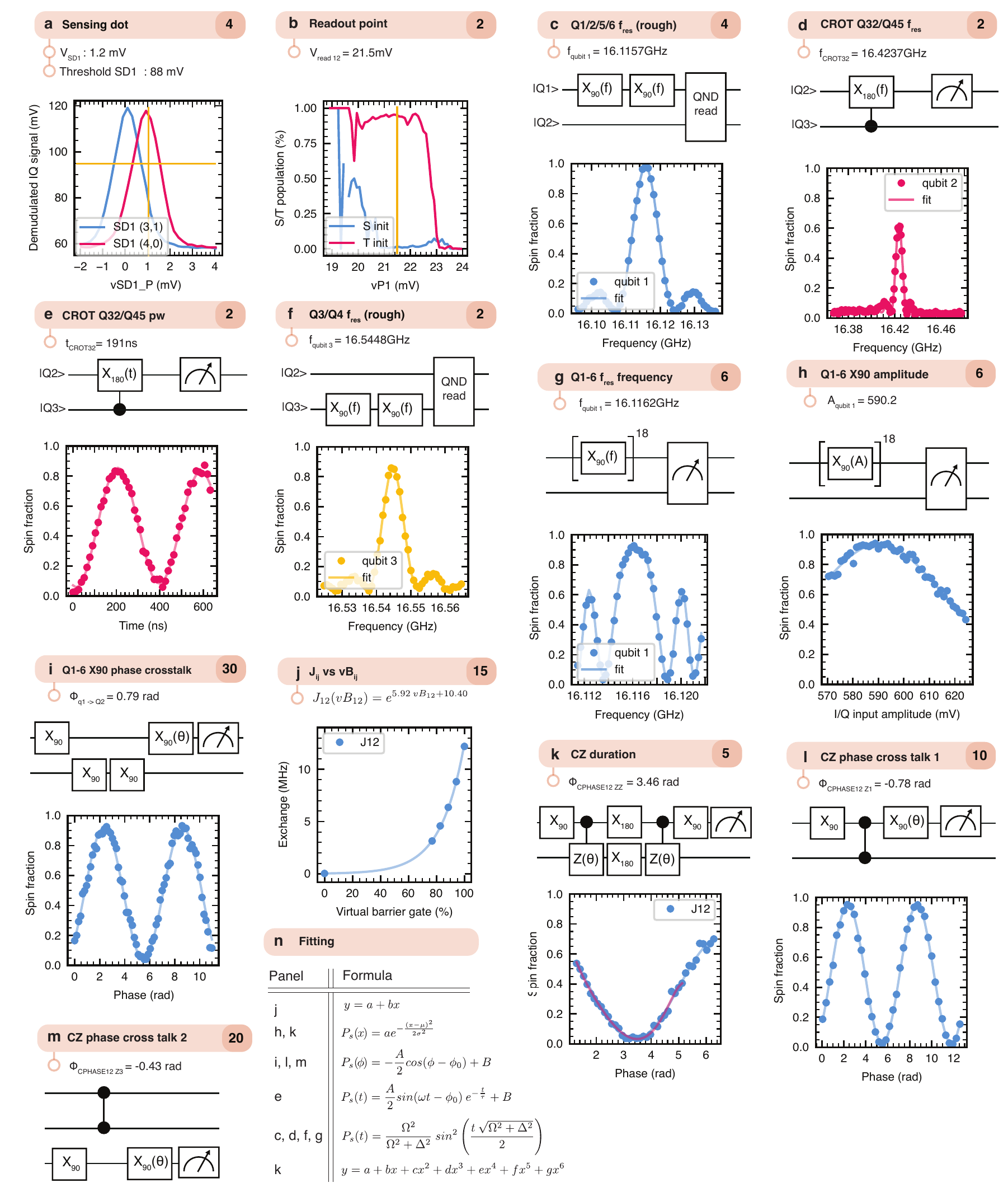}
    \captionof{figure}{\textbf{Calibrations} \textbf{a-m} Each panel shows a typical experimental dataset obtained in one of the calibration routines for the relevant experimental parameters. Every panel indicates in the header the calibration name and the corresponding number of calibration parameters. Below the header, the values extracted from the data is indicated.  See the methods section for a detailed description. \textbf{n}, Table showing the formulas used to fit the data for the indicated figure panels.}
	\label{ED:calib}
\end{center}

\begin{center}
    \includegraphics{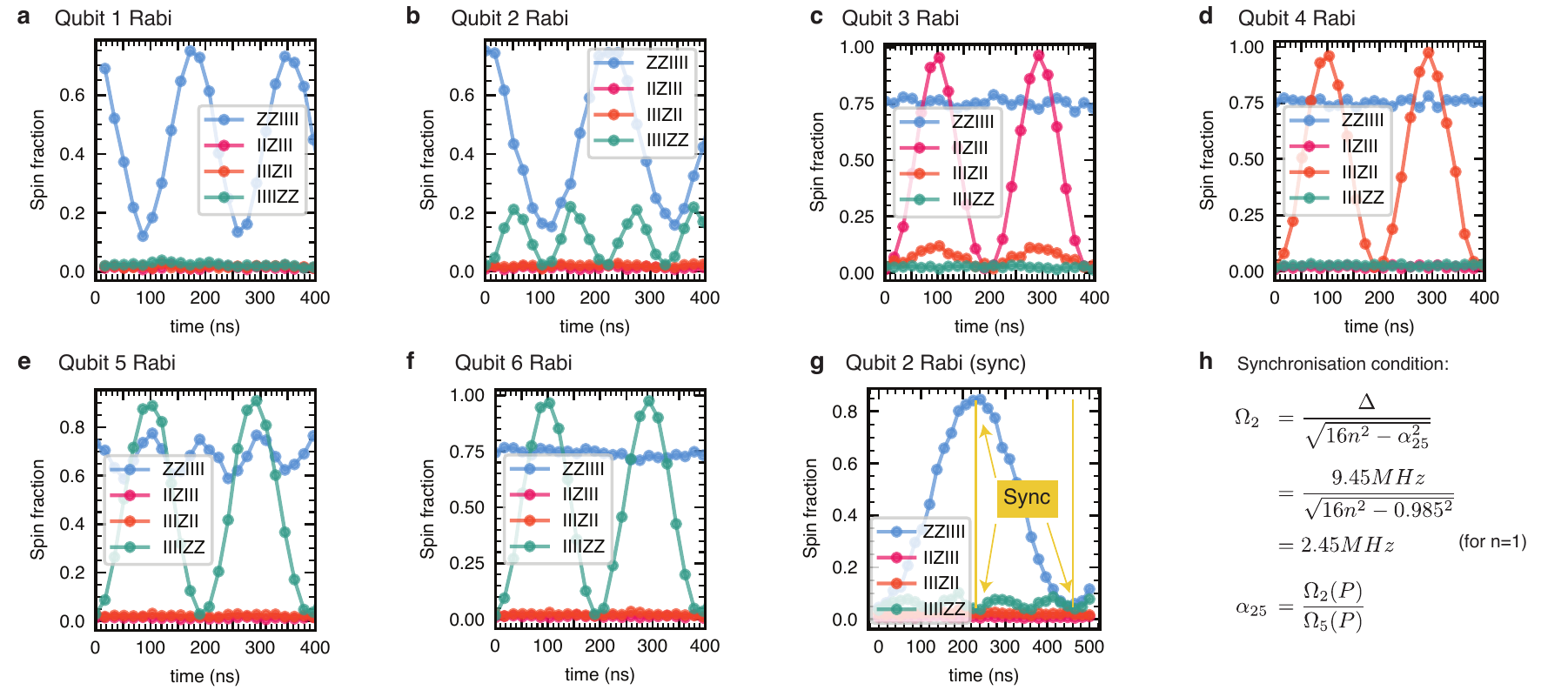}
    \captionof{figure}{\textbf{Crosstalk} 
    \textbf{a-f}, In this experiment all qubits are initialized. We observed the presence of crosstalk by measuring the expectation values of the native observables on the sample. \textbf{a-e}, Rabi oscillations executed on different qubits in each panel. Panels b, c and e display coherent oscillations of nominally idle qubits, which are a clear indication of crosstalk at the chosen Rabi frequencies. Qubits 1 and 2 show flipped measurement outcomes due to a miscalibration of sensing dot 1. This bears no influence on the conclusion from this experiment. \textbf{g}, Example of qubit 2 driven at the frequency determined by the synchronization condition. In this case the crosstalk to qubit 5 (qubit with the closest resonance frequency) was nullified by design for every multiple of a 90 degree rotation. \textbf{h}, Expressions \cite{heinz2021crosstalk} used to calculate the synchronization condition shown in panel \textbf{g}.}
	\label{ED:crosstalk}
\end{center}

\begin{center}
    \includegraphics{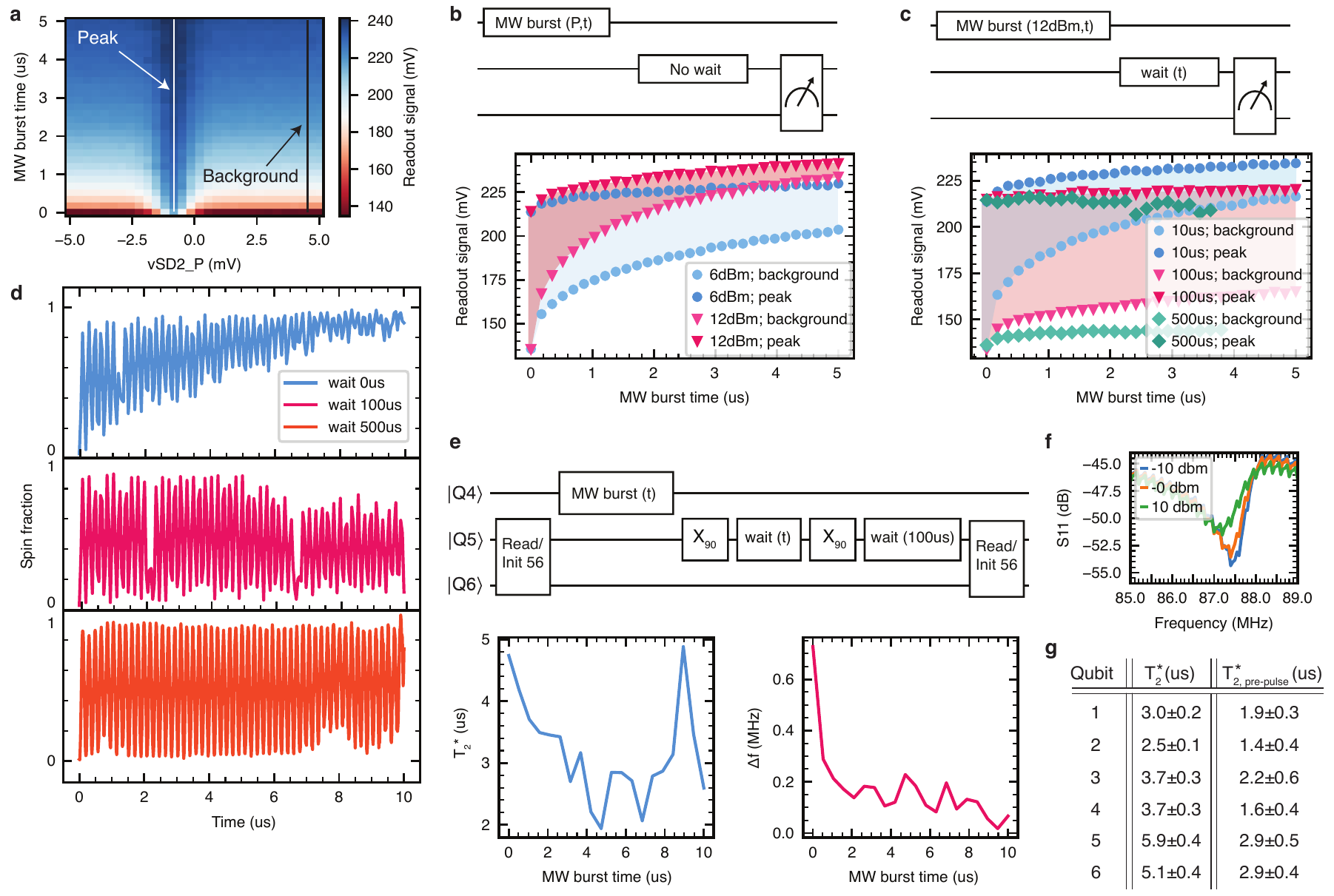}
    \captionof{figure}{\textbf{Heating effects} 
    \textbf{a}, A microwave burst of 16 GHz at 12 dBm output power and of a variable duration is applied before scanning the sensing dot virtual plunger gate. We observe no shift of the Coulomb peak, indicated by the white line, but the background signal increases significantly. Also the peak signal increases, but by a smaller amount. \textbf{b}, Linecut along the Coulomb peak and linecut parallel to the Coulomb peak of panel \textbf{a}, showing the peak and background signal as a function of the MW burst duration, for two different powers. The shaded area indicates the net signal, which is smaller the larger the applied power and the longer its duration. \textbf{c}, Variation on panel \textbf{b}, where we introduce a waiting time after the microwave burst, as indicated in the schematic. The longer the wait time, the more the original SNR is recovered, and the higher the readout fidelity will be. \textbf{d}, Rabi oscillations with different waiting times introduced before the qubit readout. The 500~\textmu s wait time allows for recording long lived Rabi oscillations, while with no wait time prior readout, the contrast vanishes as the perceived spin fraction converge towards 1, due to shifts in the sensing dot signal and background (the threshold for single-shot analysis was kept fixed). \textbf{e}, Schematic showing the circuit used to investigate the effect of a pre-pulse (labelled MW burst ($t$)) on the qubit properties. A microwave burst of 6 dBm is applied before running a Ramsey experiment. We extract the change in T$_2^*$ and Larmor frequency for qubit 5 for different microwave burst times as shown in the plots. \textbf{f} Return loss of the RF readout circuit for different powers of continued microwave driving (i.e. driving at the qubit frequencies, not for the RF readout). We observe both a shift in the RF resonance frequency and a degradation of the quality factor with higher power excitation. \textbf{g}, Extracted dephasing times T$_2^*$ with and without pre-pulse (4~\textmu s, 6 dBm), measured as illustrated in \textbf{e}.}
	\label{ED:heating}
\end{center}

\begin{center}
    \includegraphics{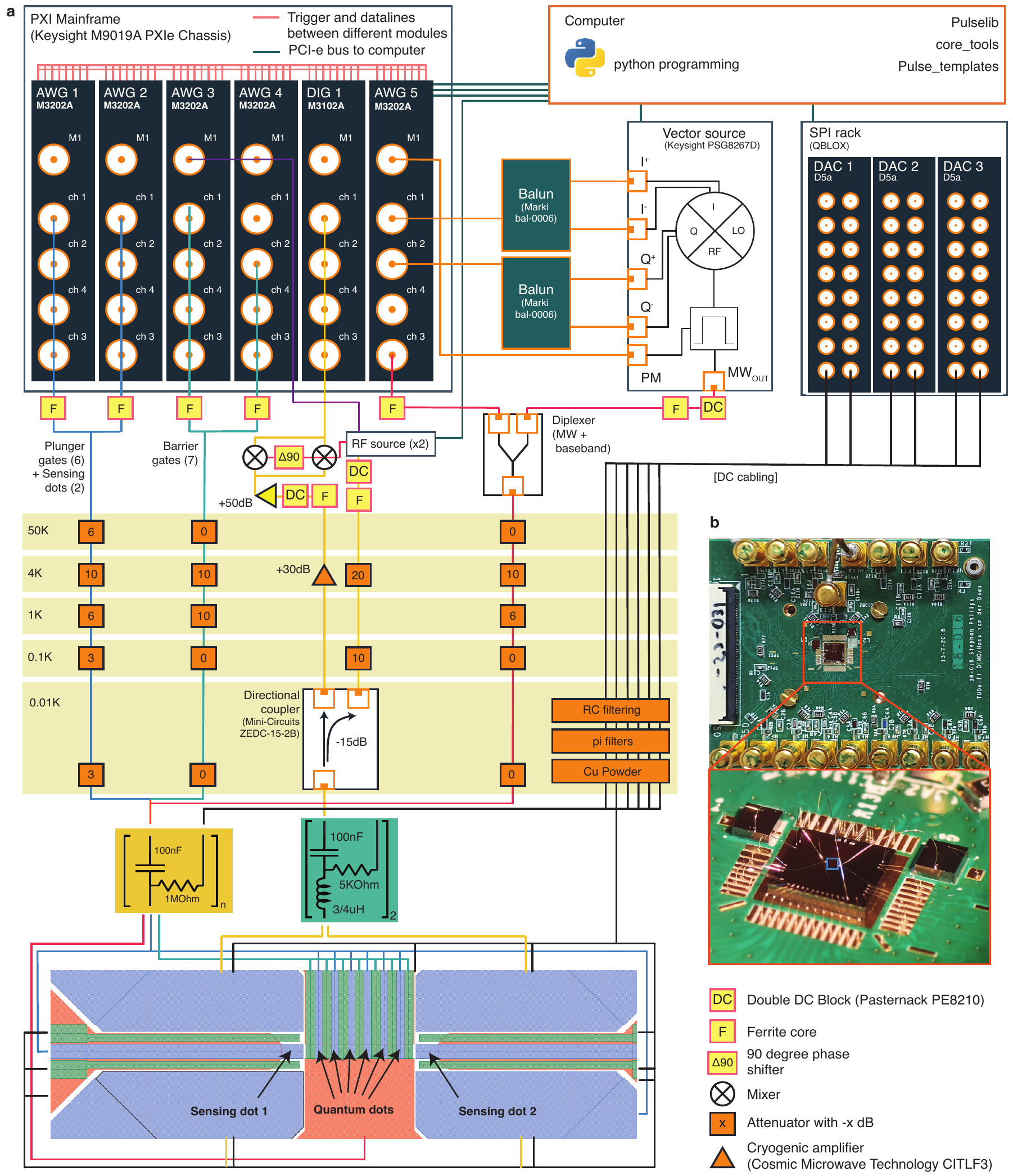}
    \captionof{figure}{\textbf{Experiment setup} \textbf{a}, Schematic overview of the experimental setup. The bulk of the experiment is controlled by Arbitrary Waveform Generators (AWG, Keysight M3202A) and digitizers (DIG , Keysight M3201A) in a PXI chassis (used for synchronization and feedback). The AWG's generate baseband pulses (0-300MHz) used for readout and two-qubit gates. These baseband signals are provided to all plunger and barrier gates of the quantum dots. Sensing dot plungers are also connected to AWG channels to allow for fully compensated virtual plungers. In addition, we use the AWG's to generate the I/Q input signals for the vector source (Keysight PSG E8267D), used to perform single-qubit gates. Using this source, a large IQ modulation bandwidth (800 MHz) can be obtained by using the differential IQ inputs. The differential signal is generated using Balun's (Marki bal-0006), which reduce the number of AWG channels and ensure excellent timing (AWG channel pairs 1,2 and 3,4 have a larger skew  ($\pm$30~ps) compared to just channel 1 and 2). We use a homemade combiner to allow for both baseband and MW control on the EDSR driving gate. Coils with ferrite cores are used to reduce low-frequency noise generated by the instruments. In addition, we use double DC blocks for any RF/MW signal used in this experiment. The schematic shows in yellow the different temperature stages of the dilution refrigerator at which the signals are attenuated and thermalized. All plunger gates have discrete attenuators in the line with a total attenuation of $\sim$ 28 dB, and barrier gates of $\sim$ 20~dB in addition to the attenuation from the coax line itself. The barriers gates have less attenuation because of the large voltage pulses needed to achieve the desired $J_{\mathrm{on}}/J_{\mathrm{off}}$ ratio. We use bias tees on the sample PCB with a RC time constant of 100~ms to combine baseband and microwave signals with a static DC voltage. We generate the carrier signal using a homemade RF source (one carrier per sensing dot), which we route into the dilution refrigerator using a combiner. A marker output channel of the AWG's is connected to the the RF sources in order to only output RF power during the readout. At the 10~mK stage, we use a directional coupler to separate the reflected signal ($S_{11}$) from the incoming RF carrier. A coplanar waveguide routing the RF signal on the sample PCB is split in two and connects to bias tees, each one going to a sensing dot. Bias tees with a low resistance are chosen as it also allows us to perform DC measurements as needed. NbTiN inductors (low $C_{\mathrm{p}}$, high Q) are wire-bonded directly to the source contacts on the sample \cite{connors2020rapid,noiri2020radio} (see Supplementary materials). When the carrier signal reflects from the sensing dots, it passes again through the directional coupler and is amplified both at the 4 Kelvin stage and at room temperature. The signal is fed through two sets of mixers (1 for each SD) to demodulate the signal to baseband. We finally sample the I and Q channels with the digitizer in the PXI chassis. On the FPGA, we average the signal for a specified amount of time and optionally convert it into a boolean value using a threshold (see Extended Date figure \ref{ED:feedback}). Besides the fast baseband/RF/MW pulses, all the gates of the sample are also connected to battery-powered DACs built in-house, which supply the DC operating voltages. These DACs are very stable voltage sources that provide an 18-bit voltage resolution over a $\pm$2~V range. \textbf{b}, Image of the PCB used to mount the sample. In the zoomed-in image in red, the qubit chip and two smaller chips with NbTiN high-kinetic inductance inductors are visible.}
    \label{ED:setup}
\end{center}

\newpage

\begin{center}
 \includegraphics{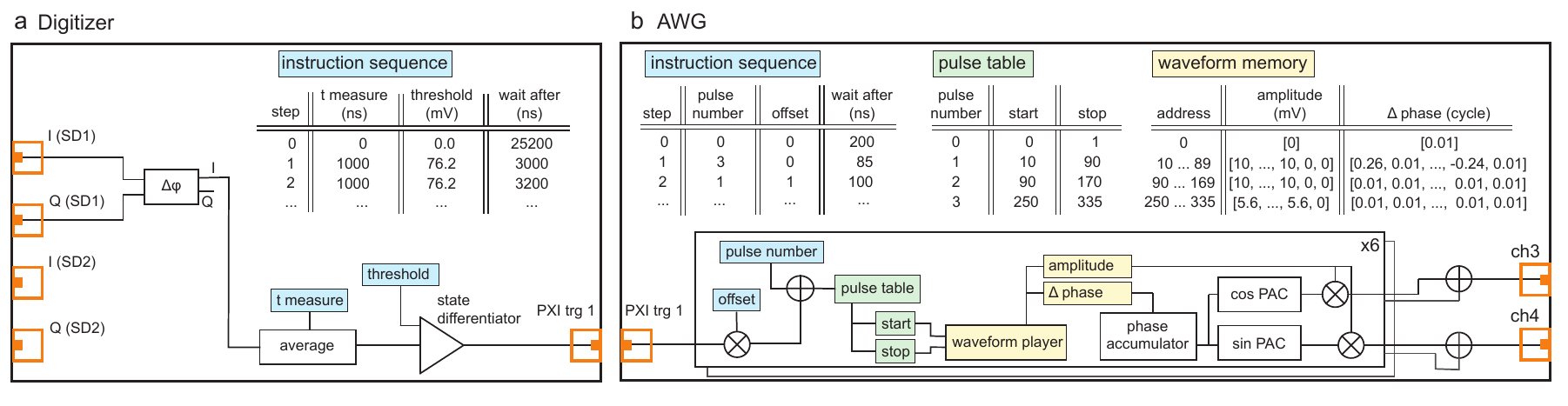}
    \captionof{figure}{\textbf{Implementation of real time feedback} 
    \textbf{a}, The sensing dot signal obtained via RF reflectometry arrives at the digitizer on two input channels (I and Q). The digitizer rotates the combined I and Q input with an angle $\Delta \phi$ and converts the vector into a scalar by dropping the Q signal. Upon a trigger from the instruction sequence, the signal is averaged for a time $t_{\mathrm{measure}}$ and compared with a threshold to infer the qubit state. The result is written both to the DRAM and the PXI trigger line. 
    \textbf{b}, We use IQ modulation to shape the MW pulses that are used for EDSR qubit control. The waveform memory stores the amplitude (envelope) and phase information for all the microwave bursts used in the experiment, as well as for the necessary single-qubit phase corrections. We upload  waveforms for microwave bursts corresponding to X$_{90}$, X$_{-90}$, Y$_{90}$ and Y$_{-90}$ rotations for each qubit. The pulse table contains the start and stop memory addresses of each control pulse present in the waveform memory. For every single-shot experiment, the AWG steps through the instruction sequence, which defines all the single-qubit gate pulses that need to be executed during the experiment. When the offset flag of an instruction in the instruction sequence is 1, the current value on the PXI trigger (0/1) is added to the pulse number that will be played from the pulse table. This bitwise addition implements real-time feedback. In the present experiment, either X$_{90}$ bursts will be executed or zero amplitude bursts, depending on the PXI trigger value. When an instruction sequence is ran, the amplitude and phase information are read from the waveform memory for the selected pulse numbers one after the other. The differential phase ($\Delta$ phase) is added for every rendered sample to the phase accumulator (which controls the qubit frequency) and is then converted to an in-phase (I) and quadrature (Q) signal by the phase-to-amplitude converters (PAC). These I and Q signals are multiplied with the amplitude envelope of the waveform and are then passed to the outputs of the AWG and from there to the vector source. We can run up to twelve sequencers in parallel in a single AWG. In this case 6 sequencers are used, one for every qubit.
    }
	\label{ED:feedback}
\end{center}

\begin{center}
    \begin{tabular}{c||c|c|c|c|c|c|c}
	 & $\mathrm{vB0}$   & $\mathrm{vB1}$ & $\mathrm{vB2}$ & $\mathrm{vB3}$ & $\mathrm{vB4}$ & $\mathrm{vB5}$ & $\mathrm{vB6}$ \\
	\hline\hline
	${V_{J_{12}}}$ & -1   & 1 & -1.6 & 0 & 0 & 0 & 0 \\
	${V_{J_{23}}}$ & 0   & -0.2 & 1 & -0.5 & 0 & 0   & 0 \\
	${V_{J_{34}}}$ & 0   & 0 & -0.3 & 1 & -0.3 & 0  & 0 \\
	${V_{J_{45}}}$ & 0   & 0 & 0 & -0.9 & 1 & -0.9  & 0 \\
	${V_{J_{56}}}$ & 0   & 0 & 0 & -0.2 & -0.9 & 1  & 0 \\

\end{tabular}
    \captionof{table}{\textbf{Exchange pulses.} In order to achieve sufficiently high $J_{\mathrm{on}}/J_{\mathrm{off}}$, we use a combination of barrier gate pulses, where we pulse a barrier gate in between the target qubit pair to a more positive voltage and at the same time we pulse the voltage on the barrier gates on the outer side of target qubit pair to a less positive voltage. This pushes the two quantum dots towards each other further enhancing the tunnel coupling.}
	\label{ED:Jpulses}
\end{center}



\section*{Supplementary material}

# SUPPLEMENTARY MATERIALS
## Universal control of a six qubit quantum processor in silicon

Stephan G.J. Philips[*1], Mateusz T. Mądzik[*1], Sergey V. Amitonov[1], Sander L. de Snoo[1], Maximilian Russ[1], Nima Kalhor[1], Christian Volk[1], William I.L. Lawrie[1], Delphine Brousse[2], Larysa Tryputen[2], Brian Paquelet Wütz[1], Amir Sammak[2], Menno Veldhorst[1], Giordano Scappucci[1], and Lieven M.K. Vandersypen[†1]

[1] QuTech and the Kavli Institute of Nanoscience, Delft University of Technology, 2600 GA Delft, The Netherlands.
[2] QuTech and Netherlands Organization for Applied Scientific Research (TNO), Delft, The Netherlands.

February 18, 2022

# Contents

---
[*]These authors contributed equally
[†]To whom correspondence should be addressed; E-mail: L.M.K.Vandersypen@tudelft.nl

# 1 Micromagnet design

Most micromagnets used in spin qubit experiments have been designed to support up to 2 to 3 qubits [1, 2, 3]. In this experiment, a suitable gradient was needed across a six-qubit device. A schematic of the micromagnet design, along with the coordinate system is shown in Supplementary Fig. 1. When designing the magnet, we set the following requirements:

- Target Rabi frequency of 5-10 MHz. The higher the driving speed, the more operations can be completed within the coherence time. We limit the target frequency to 10 MHz as we have found the Rabi frequency to become non-linear in the driving amplitude for higher Rabi frequencies (see also ref. [4]). Based on prior experience, this translates to a target gradient in the transverse component of the micromagnet stray field of at least 1 mT/nm.

- All the qubits are driven by a single gate using frequency multiplexing. Overlap in resonance frequencies must thus be avoided. We target a frequency difference of 100 MHz between different qubits when the micromagnets are fully magnetized. This value is chosen as it results in an acceptable level of crosstalk when driving the qubits at a Rabi frequency of 5-10 MHz.

- Following [4], we aim at a maximal decoherence gradient ($|\frac{\partial B_z}{\partial z}| + |\frac{\partial B_z}{\partial y}|$) of 0.1 mT/nm. In this way, qubit coherence is not impacted assuming an intrinsic qubit linewidth of $\delta f \sim 10$ kHz. We report the decoherence gradient calculated at the center of the quantum dot location.

To design the micromagnets, we parametrized a model of the magnets and optimized a number of design parameters (e.g. the magnet-magnet separation, height of the magnets, size and angle of the slanting part and distance from the quantum well). The core of the calculations was based on the python package magpylib [5], which allows for fast analytical solutions of simple magnet geometries (our wrapper can be found at [6]). We assume full magnetization of the micromagnet in our simulations. The result of the optimization is shown in Supplementary Fig. 1 b-f. In panel b-c, one can visually inspect the micromagnet stray magnetic field profile. Panel d-f show line cuts of magnetic properties described above, along the length of the six-qubit array. We are able to satisfy all our design targets.

Supplementary Fig. 1g-i show the measured frequency distributions of the qubits on three samples from different fabrication runs. The measured frequency profile for sample A is comparable to the simulated profile of Supplementary Fig. 1d. For samples B and C, the measured frequencies are very different than expected based on the simulations. The parabolic trend seen in Sample C (the sample discussed in the main text paper) can be roughly reproduced by assuming that the magnet boundary is displaced. The solid white lines in Supplementary Fig. 1j show the bottom and the top of the magnet, indicating that the micromagnet sidewalls are slightly tilted, which indeed effectively displaces the magnet boundary. When simulating the frequency profile with the black dashed line as the magnet boundary (as a rough approximation), we obtain the profile shown in purple in Supplementary Fig. 1i. The qubit frequencies reported in these plots are reproducible between different cooldowns of the same device.

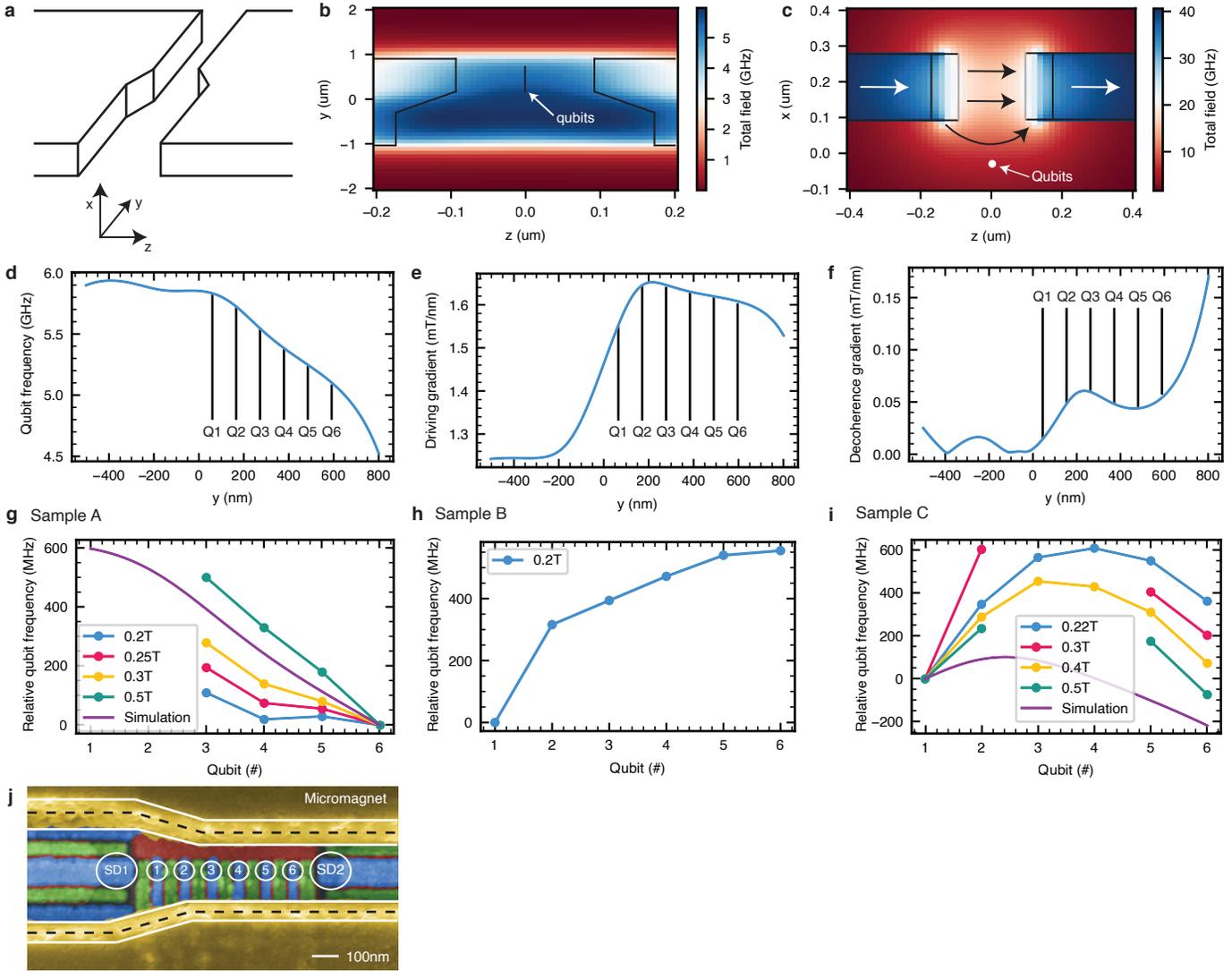

**Supplementary Figure 1 | Micromagnet simulations and experimental values a**, Sketch of the micromagnet structure in 3D. The micromagnets are magnetized along the z direction, with a magnetization vector of $M = (0\ T, 0\ T, 1.5\ T)$ as estimated from the experiments. The maximum magnetisation is reached in the experiment when 0.1-0.2T external field is applied to the magnets. **b**, Simulation of the total field generated by the micromagnets in the z,y plane, where the quantum dots reside. A projection of the micromagnets is shown by the black lines overlaid in the figure. **c**, Simulation of the total field generated by the micromagnets in the z,x plane. **d**, Simulated values for the qubit frequencies (without the contribution of the externally applied magnetic field) along the y axis, with the target qubit positions indicated. **e**, Simulated values for the (transverse) driving gradient of the micromagnet ($|\frac{\partial B_x}{\partial z}| + |\frac{\partial B_y}{\partial z}|$) along the y axis, with the target qubit positions indicated. **f**, Simulated values of the decoherence gradient of the micromagnet ($|\frac{\partial B_z}{\partial z}| + |\frac{\partial B_z}{\partial y}|$) along the y axis, with the target qubit positions indicated. **g-i** Experimentally measured qubit frequencies for three different samples with the same micromagnet design. All frequencies in the plots are taken in reference to one qubit, to clearly display the difference in qubit frequency distribution versus the different fields applied to the sample. The simulated curve in panel **i** is computed assuming the magnet boundary is as indicated by the dashed black line in panel **j**. **j**, False colored SEM image of a sample similar to sample C and fabricated in the same fabrication run. The white lines indicates the top and bottom edge of the micromagnet, with inner lines indicating the intended step boundary. The black, dashed line is the average distance between top and bottom edge of the magnets.

3

# 2 Different samples tested for this experiment

|  | Device A | Device B | Device C |
|---|---|---|---|
|  | **Design** | | |
| Number of qubits | 6 | 6 | 6 |
| Number of sensing dots | 3 | 2 | 2 |
| Access to reservoir from dots | 1, 6 | 1, 3, 4, 6 | 1, 6 |
| Dot pitch | 100 nm | 80 nm | 90 nm |
| Tuning for sufficient $t_\mathrm{c}$ | 3,1,3,1,3,1 | 1,1,1,1,1,1 | 1,1,1,1,1,1 |
| Comment | Right SD unusable due to faulty source contact | | |
|  | **Valley splitting** (µeV) | | |
| Dot 1 | 160 | 118.9 | 220 |
| Dot 2 | 130 | 160.8 | 140 |
| Dot 3 | 0 | 113.4 | 105 |
| Dot 4 | 173 | 160.9 | 138 |
| Dot 5 | – | 56.8 | 220 |
| Dot 6 | – | 148.4 | 300 |
| Comment | Valley splitting in dot 3 too low for qubit experiments | | |
|  | **Dephasing time $T_2^*$** (µs) | | |
| Dot 1 | 12.7 | 3.0 | 3.0 |
| Dot 2 | 5.5 | 3.2 | 2.5 |
| Dot 3 | 3.4 | 6.0 | 3.7 |
| Dot 4 | 7.2 | 3.0 | 3.7 |
| Dot 5 | – | 6.5 | 5.9 |
| Dot 6 | – | 6.4 | 5.1 |
|  | **Hahn echo decay time $T_2^\mathrm{h}$** (µs) | | |
| Dot 1 | 24.8 | 77 | 14.0 |
| Dot 2 | 28.1 | 47 | 21.1 |
| Dot 3 | – | 75 | 40.1 |
| Dot 4 | 26.9 | 48 | 37.2 |
| Dot 5 | – | 58 | 44.7 |
| Dot 6 | – | 41 | 26.7 |
|  | **General comment** | | |
|  | Sample discarded due to low valley splitting in dot 3 | Sample discarded due to low EDSR drive quality | Sample used in this experiment |

**Supplementary Data Table 1 | Properties of different samples tested for this experiment.** In the course of this experiment, we have modified the sample design to match the requirements set for 6-qubit control. Device A has been discarded due to the low valley splitting in dot 3. Additionally, we could only use 4 out of 6 quantum dots, due to a failure on the right sensing dot. Due to the 100 nm dot pitch, it was necessary to tune the device to the (3,1,3,1,3,1) charge configuration in order to achieve sufficient tunnel coupling $t_\mathrm{c}$ between adjacent dots. Device B allowed for 6-qubit operation, however, the EDSR drive quality was poor, with Rabi oscillations decaying within a few periods. We have used Device C with a 90 nm pitch for the data presented in the main text. The reported valley splittings are measured using magnetospectroscopy. Since there are no reservoirs available in the middle of the device, we measure the valley splitting by performing spectroscopy on the anticrossing between the (2,0) and (1,1) at a relatively low tunnel coupling [7].

# 3 Coherence times and visibilities

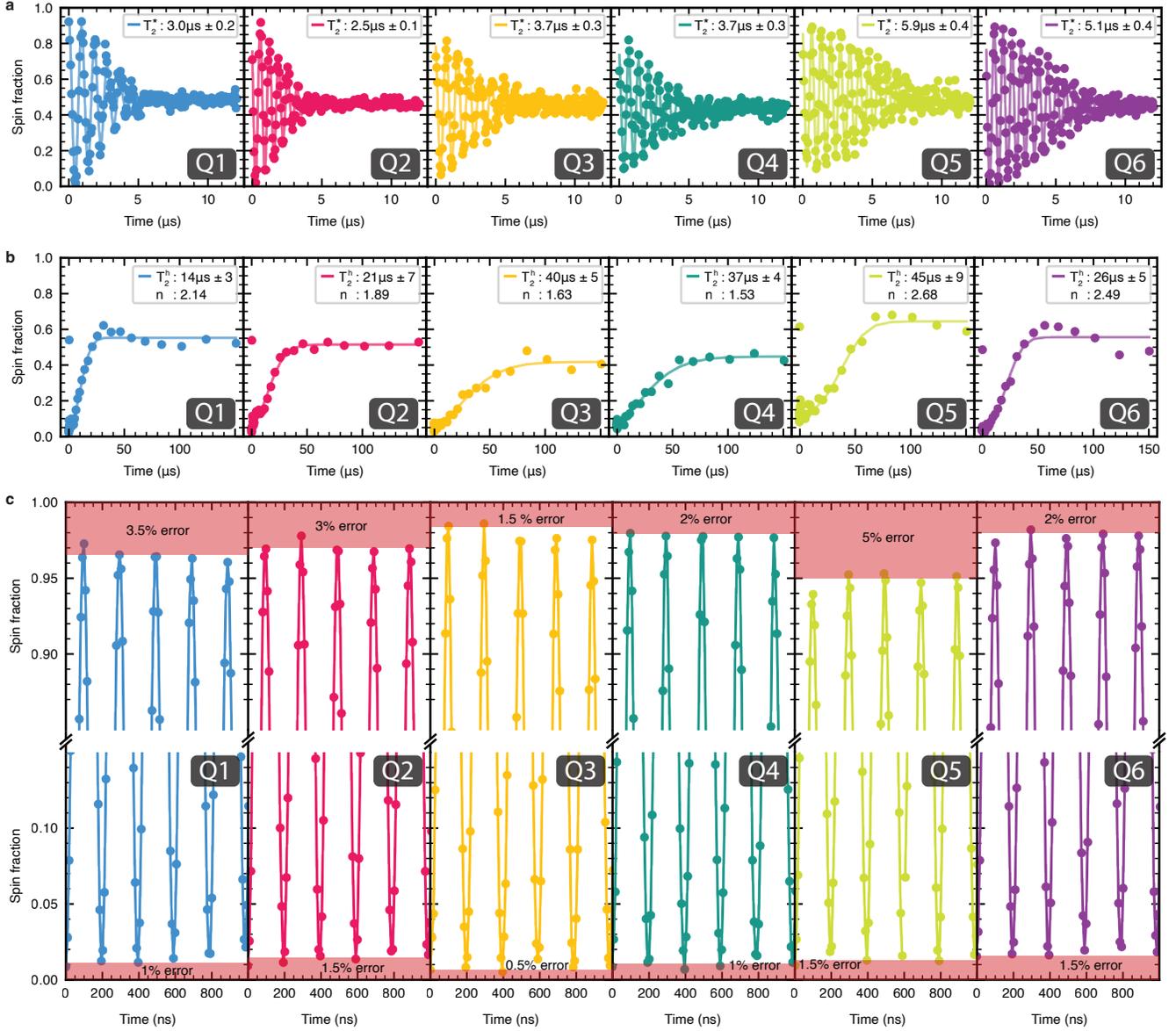

**Supplementary Figure 2 | $T_2^*$, $T_2^h$ and visibility measurements a**, $T_2^*$ measurement for each qubit. These measurements were measured without any pre-pulse and are fitted to a Gaussian decay: $P_s(t) = A\cos(\omega t + \phi)\exp\left(-\frac{t^2}{T_2^{*2}}\right) + B$.
**b** $T_2^h$ measurement for each qubit. These measurements are also performed without any pre-pulse and are fitted using: $P_s(t) = A\exp\left(-\frac{t^n}{T_2^{h\,n}}\right) + B$. **c**, Detailed plots of the Rabi oscillations shown in the main text. The errors contributing to loss of visibility are estimated by eye.

# 4 Calibration log

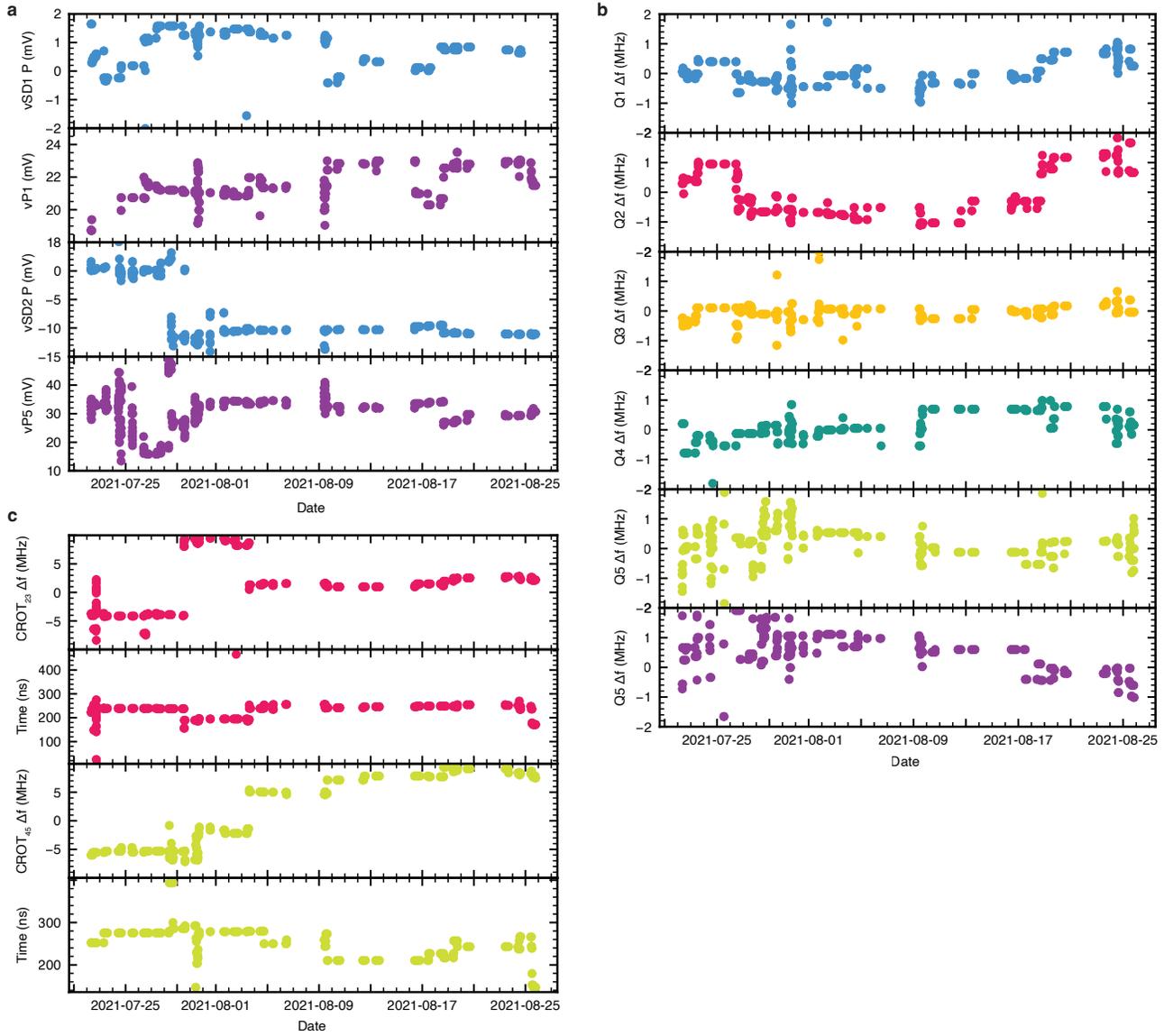

**Supplementary Figure 3 | Calibration log** of the sample during a period of approximately one month. Every data point in the calibration plot represents a result of a calibration measurement **a**, Calibration results of the sensing dot and readout point. The panels with the blue dots show the calibration results of the optimal operating point of SD1 (top) and SD2 (bottom). The purple dots show the readout point used for the parity readout for dot pair 12 (top) and 56 (bottom). **b**, Calibration results for the qubit resonance frequency, one plot per qubit. We plotted the deviation from the average resonance frequency. **c**, Controlled rotation calibration. The calibrated values for the driving frequency and burst duration for $CROT_{23}$ and $CROT_{45}$ are shown.

# 5 Sample design

A CAD image of the device design in shown in Supplementary Fig. 4a,b. The samples are designed without a physical mesa. A mesa is often used to prevent leakage at the bondpads where damage by a wirebonder could create a contact to the 2DEG residing below. Instead, we deposit a 200 nm silicon nitride layer underneath the bondpads to prevent leakage to the 2DEG below (Supplementary Fig. 4c). In addition, all the gates in the plunger and barrier gate layer run over a screening gate layer, blocking any current flow between the bondpad and the center of the device, as long as the screening gates are not accumulated (Supplementary Fig. 4d). As a last measure, part of the gate fan-out wiring is reduced to patterned nanowires (Supplementary Fig. 4e), which need higher voltages to accumulate.

We designed an on-chip coplanar waveguide with a characteristic impedance of 50 Ohm to optimize the power delivery of the microwave excitation. For the RF readout, we use a low capacitance design where the readout signal is applied via the source contact [8, 9, 10]. We ensure low parasitic capacitance by placing the source contact close to the center of the device ((Supplementary Fig. 4a)) and wire bond directly to these source contacts. Furthermore, running the accumulation gates over a screening gate prevents the creation of a large capacitance between accumulated 2DEG and the accumulation gate, which helps achieve a good RF readout.

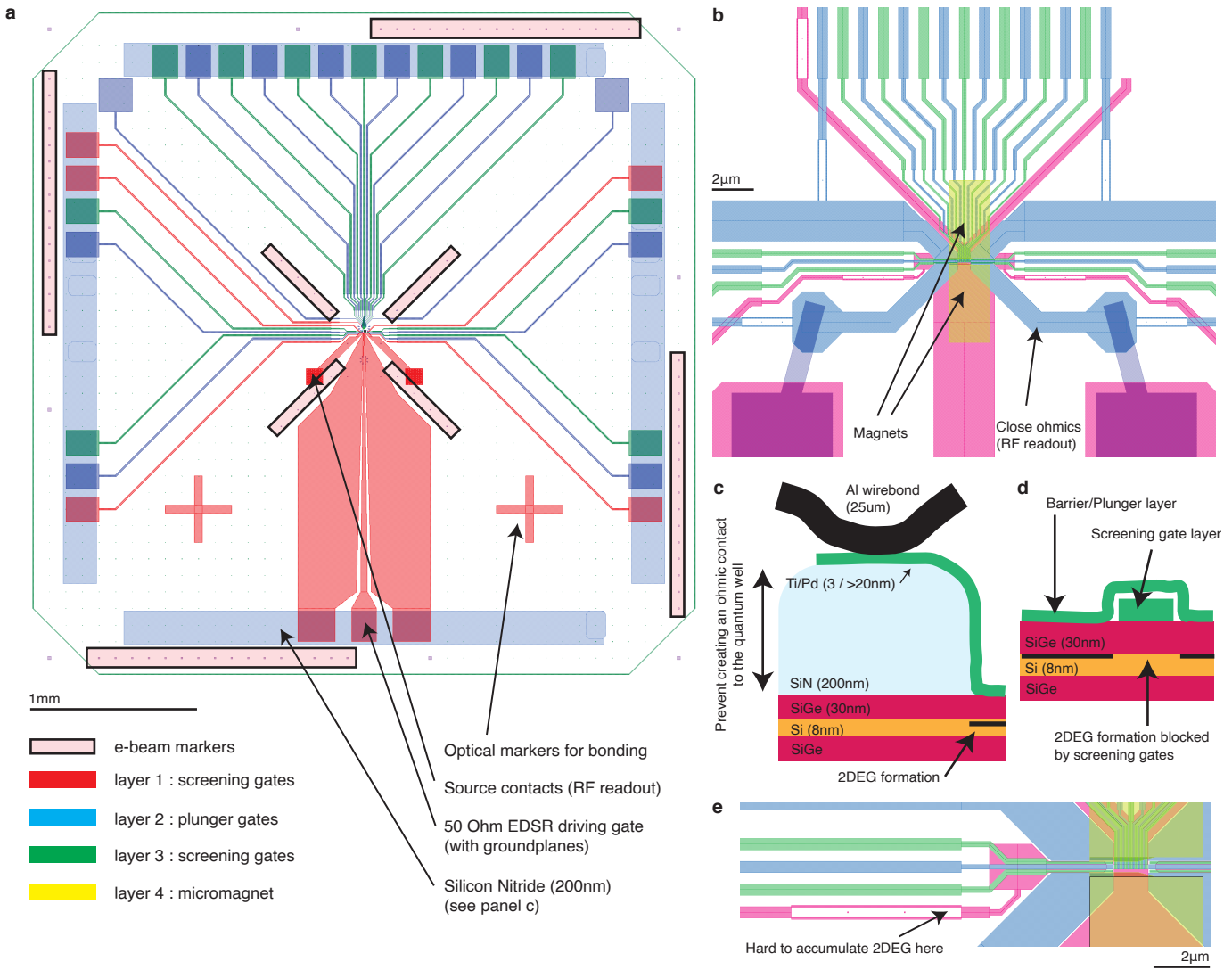

**Supplementary Figure 4 | Sample design a**, CAD image of the sample layout. **b**, Zoom in of **a**, showing the active region of the device. **c**, Silicon nitride (SiN) below the bondpad prevents punch-through of the bondwire to the quantum well. **d**, Screening gates block current flow below the gates between the bondpads and the active device area. **e**, Narrow segments of the gates increase the accumulation voltage underneath, and thereby also block current flow underneath.

# 6 Entanglement Witnesses

Entanglement witnesses are an efficient means to check whether a state is entangled. A witness is constructed in such a manner that its expectation value is negative in case the state is entangled. When choosing an entanglement witness, usually one needs to balance the number of operators measured versus the possible states that can be detected. Examples for GHZ states include the optimal, stabilizer and Mermin witness [11]. We choose a witness which requires a few more measurements but is able to detect entanglement across a larger part of the space with entangled states:

$$W = III - |\psi_{\text{GHZ}}\rangle \langle \psi_{\text{GHZ}}| . \tag{1}$$

This operator can be decomposed into Pauli operators. For three qubits, this results in:

$$\begin{aligned} W = &\frac{3}{8} \langle III \rangle - \frac{1}{8} \langle IZZ \rangle - \frac{1}{8} \langle XXX \rangle + \frac{1}{8} \langle XYY \rangle + \\ &\frac{1}{8} \langle YXY \rangle + \frac{1}{8} \langle YYX \rangle - \frac{1}{8} \langle ZIZ \rangle - \frac{1}{8} \langle ZZI \rangle . \end{aligned} \tag{2}$$

We use similar procedures as for the state tomography to remove the SPAM error of the witness operators.

# 7 Quantum state tomography

Quantum state tomography is used to obtain the density matrices for the qubits. We used the maximum likelihood method to obtain the density matrices reported in this paper.

The concept of quantum state tomography is based on the idea that every density matrix can be decomposed in a set of orthogonal basis states (e.g. Pauli basis):

$$\rho = |\psi\rangle \langle \psi| = \frac{1}{2^N} \sum_i^{4^N} \langle \psi| \hat{V}_i |\psi\rangle \cdot \hat{V}_i , \tag{3}$$

where $\hat{V}_i$ is the i$^{th}$ basis state. Here $\hat{V}_i$ are of the form $\{I, X, Y, Z\}^{\otimes N}$, and N is the number of qubits. In the experiment we measure the expectation value $M_i$ of all the possible $\hat{V}_i$'s; $M_i = \langle \psi| V_i |\psi\rangle$ and reconstruct a first estimate of the density matrix using equation 3. To ensure that the obtained density matrix $\rho$ is valid, we take the closest positive semi-definite matrix and ensure that the norm of $\rho$ is one.

The resulting state can be used as input for the Maximum Likelihood Estimation (MLE) [12]. This is procedure to obtain the most likely state for a given set of $M_i$, using an optimizer for the following cost function:

$$\sum_i^{4^N} |M_i - tr(\rho V_i)| \tag{4}$$

To make this method work well, we use so called $T$ matrices, which is the lower triangular matrix, where every non-zero entry is a variable. The relation to the density matrix is $\rho = TT^\dagger$. An initial guess of $T$ is obtained by performing a Cholesky decomposition on the density matrix used as input for MLE.

The fidelity of the estimated density matrix can be further increased by taking into account the readout errors of the system. When measuring an observable, the following matrix can be used to describe the measurement process:

$$P_{\text{meas}} = S_k P_{\text{real}} = \begin{pmatrix} F_{k,-1} & 1 - F_{k,+1} \\ 1 - F_{k,-1} & F_{k,+1} \end{pmatrix} P_{\text{real}} \tag{5}$$

where $P_{\text{meas}}$ are the measured spin spin probabilities for the k$^{th}$ observable (while we are interested in the real probability amplitudes). $P_{\text{real}} \dot{F}_{k,-1}$ ($F_{k,+1}$) is the corrected for readout error probability to obtain $-1$ ($+1$) as measurement outcome for the $k^{th}$ observable. These numbers are derived from the visibility of single-qubit Rabi oscillations such as in main text Fig. 2, but taken with the same initialization and readout sequences as are used in the circuits for preparing the density matrix.

The following observables are measured:

| k | observable |
|---|---|
| 1 | $ZIIIII$ |
| 2 | $IZIIII$ |
| 3 | $ZZIIII$ |
| 4 | $IIZIII$ |
| 5 | $IIIZII$ |
| 6 | $IIIIZI$ |
| 7 | $IIIIIZ$ |
| 8 | $IIIIZZ$ |

and have their own associated $S_k$ matrix. We decompose other measurement operators we want to measure into elements of this set, for example $ZZZIII$ is decomposed into a combination of the $ZZIIII$ and $IIZIII$ operators, which we can simultaneously measure. The procedure we would use to generate to expectation value would be give by :

$$\langle ZZZIII \rangle = \text{Tr}(ZZ \cdot (S_3 \otimes S_4)^{-1} \cdot P_{\text{meas}}) \tag{6}$$

Where $P_{\text{meas}}$ is a vector with the probabilities of the four possible measurement outcomes.

A comment needs to be made on the validity of this approach, as the SPAM errors are most likely spread over initialization and readout. The above method of removing readout errors could then artificially result in faulty results (e.g. expectation values above 1) in specific circumstances. As an example, let us assume that initialization and readout are error-prone for one qubit (Q1) and perfect for the other qubit (Q2). Characterization of SPAM errors would give us a matrix $S_1$ with which to correct the measurement outcomes for Q1.

Now consider performing state tomography after running a CNOT or SWAP operation:

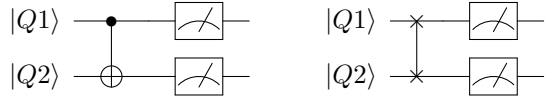

In the first case, correction using the matrix $S_1$ will accurately remove the SPAM errors from the measured values for Q1. For qubit two, errors have been propagated from Q1, but these errors are not removed by the analysis as there are no SPAM errors on Q2 by itself. In this case, SPAM error removal thus only removes a subset of the errors introduced by SPAM. In the second case, we get for both qubits the wrong results. For Q1 the corrected expectation values could exceed one (which would be non-physical), whereas for Q2 there is no correction even though initialization errors on Q1 have propagated to Q2. In the experiments that are performed in this work, the circuits focus on entangling rather than swapping states. For that reason we believe that we do not introduce nonphysical elements by applying the SPAM correction; rather it cannot remove all errors and will tend to worsen the actual fidelity.

## 8 Error channels in the experimental data: dephasing and "heating"

In order to examine the impact of dephasing on the measured density matrices of main text Figs. 4-5, we compare with the results of two numerical simulations performed according to the following methodology.

### 8.1 State tomography simulations

The multi-qubit system is well approximated by the Hamiltonian

$$H = H_{\text{Zeeman}} + H_{\text{Heisenberg}}, \tag{7}$$

consisting of the Zeeman interaction

$$H_{\text{Zeeman}} = \sum_{j=1}^{6} \frac{g_j \mu_B}{2} \boldsymbol{B}_j \cdot \boldsymbol{\sigma}_j \tag{8}$$

and the isotropic Heisenberg exchange interaction

$$H_{\text{Heisenberg}} = \sum_{\langle j,k \rangle} \frac{J_{j,k}}{4} \boldsymbol{\sigma}_j \cdot \boldsymbol{\sigma}_k. \tag{9}$$

Here, $\boldsymbol{\sigma}_j$ is the vector of the Pauli matrices acting on qubit $j$, $\boldsymbol{B}_j = (v_{ac,j}(t)\partial B_{x,j}\cos(2\pi f_j), 0, B_{z,j})^T$ is the combined magnetic field, $\partial B_{x,j}$ the micromagnet gradient orthogonal to the qubit array axis (expressed in units of Tesla/V), $v_{ac,j}$ the voltage amplitude applied for driving EDSR, and $J_{j,k} = J_{\text{res},j,k}\exp(2\alpha_{j,k}v_{B,j,k}(t))$ is the exchange interaction between qubit $j$ and $k$, where $J_{\text{res},j,k}$ is the residual exchange for $v_{B,j,k}(t) = 0$. The sum $\langle j,k\rangle$ runs over all neighboring pairs in the linear array.

For the numerical simulations, we solve the time-dependent Schrödinger equation

$$i\hbar\frac{d}{dt}|\psi(t)\rangle = H|\psi(t)\rangle \tag{10}$$

by discretizing $H(t + \Delta t)$ into segments of length $\Delta t$ taking $H(t)$ constant in the time-interval $[t, t + \Delta t]$. We compute the unitary propagator according to

$$U(t + \Delta t) = e^{-\frac{i}{\hbar}H(t+\Delta t)}U(t), \tag{11}$$

where $\hbar = h/(2\pi)$ is the reduced Planck constant. The simulations are performed in the multiply rotating frame which removes the Larmor precession of each qubit around the average external magnetic field $\sum_{j=1}^{N}B_{z,j}/N$. By making the so-called rotating wave approximation (RWA) we neglect counter-rotating terms such that we can choose $\Delta t = 10$ ps as a sufficiently small time step.

The simulated GHZ states are the final states $\rho_f = U_f\rho_{\text{init}}U_f^\dagger$ obtained by applying the pulse sequence of main text Fig. 5a to the ideal input state $\rho_{\text{init}}$ (here $U_f$ is shorthand for the product of the successively applied unitary operations). For simplicity each GHZ simulation is performed on the subspace of only the involved qubits, motivated by the small residual exchange to the other qubits.

Low-frequency noise is included in the simulation via quasistatic fluctuations $H_{\text{Zeeman}} \to H_{\text{Zeeman}} + \sum \xi_J\sigma_{z,j}$ and averaging the final result over 5000 random initializations. Here $\xi_J$ are Gaussian stochastic variables with mean $\langle\xi_J\rangle = 0$ and variance $\langle\xi_J^2\rangle = h^2/2(\pi\text{T}_{2,j}^*)^2$.

## 8.2 Discussion

In the tables below we present a comparison of the experimental results (RAW data: without SPAM error removal; Processed data: with SPAM error removal) to the outcomes of the two simulations with different input parameters. The first simulation uses the $\text{T}_2^*$ values from the table in main text Fig. 2e. The second simulation takes the $\text{T}_2^*$ values obtained after a 4 µs microwave bursts applied off-resonance before the pulse sequence used to measure $\text{T}_2^*$. We find systematically that such a prepulse causes both a shift in the qubit frequency and a reduction in $\text{T}_2^*$, by amounts that depend the microwave burst duration and power (see Extended Data Fig. 7). In the actual experiments, prepulses shorter than 4 µs are used, to strike a balance between saturating the frequency shift and not reducing $\text{T}_2^*$ too much. We can thus expect that dephasing reduces the off-diagonal elements in the measured density matrices of Fig. 4-5 by an amount that lies in between the case of the two simulations, with the shorter and longer sets of $\text{T}_2^*$ values. Furthermore, the longer the state preparation sequence, the shorter the resulting $\text{T}_2^*$, which is consistent with the fact that the off-diagonal entries are lower for the GHZ states of qubits 2-4 and 3-5 than for GHZ states of qubits 1-3 or 4-6.

The diagonal entries deviate most from the ideal expectation for the GHZ states involving qubits 2-4 and 3-5. State preparation of qubits 1-3 or qubits 4-6 by itself works very well, but we find consistently that state preparation of qubits 4-6 is somewhat degraded when it follows state preparation of qubits 1-3 (see Extended Data Fig. 3), and vice versa. In the simulations we assume perfect initial states in order to probe only the effects of dephasing.

Finally, the qubit frequency shifts from the prepulses and the pulses applied during state preparation lead to reproducible phase shifts seen in the experimentally prepared GHZ states, deviating from the ideal GHZ state $(|000\rangle + |111\rangle)/\sqrt{2}$.

The frequency shifts and reduced dephasing times generally bear signatures of heating, as in previous reports [13, 14]. Their microscopic origin is at present not well understood.

Generally, we find that the reduced coherence times in the second simulation provides us with state fidelities closely resembling the experimental data after SPAM removal. We neglect the non-ideal phase component in the experimental data when making this comparison, as we have omitted frequency shits and virtual-Z gates in our simulation.

## Qubits 12

| | Density Matrices | | | | Fidelity |
|---|---|---|---|---|---|
| Raw data | $\begin{pmatrix} 0.46e^{0.0i} & 0.047e^{-1.989i} & 0.063e^{1.827i} & 0.414e^{-0.0i} \\ 0.047e^{1.989i} & 0.034e^{0.0i} & 0.02e^{-0.016i} & 0.018e^{-2.087i} \\ 0.063e^{-1.827i} & 0.02e^{0.016i} & 0.03e^{0.0i} & 0.104e^{-1.708i} \\ 0.414e^{0.0i} & 0.018e^{2.087i} & 0.104e^{1.708i} & 0.476e^{0.0i} \end{pmatrix}$ | | | | 88.2% |
| Processed data | $\begin{pmatrix} 0.461e^{0.0i} & 0.046e^{-2.008i} & 0.07e^{1.814i} & 0.424e^{0.001i} \\ 0.046e^{2.008i} & 0.03e^{0.0i} & 0.016e^{-0.052i} & 0.013e^{-2.489i} \\ 0.07e^{-1.814i} & 0.016e^{0.052i} & 0.03e^{0.0i} & 0.106e^{-1.721i} \\ 0.424e^{-0.001i} & 0.013e^{2.489i} & 0.106e^{1.721i} & 0.479e^{0.0i} \end{pmatrix}$ | | | | 89.4% |
| Full $T_2^*$ simulation | $\begin{pmatrix} 0.502e^{0.0i} & 0.004e^{1.455i} & 0.017e^{-2.195i} & 0.468e^{-0.096i} \\ 0.004e^{-1.455i} & 0.004e^{-0.0i} & 0.003e^{0.095i} & 0.005e^{-1.923i} \\ 0.017e^{2.195i} & 0.003e^{-0.095i} & 0.004e^{0.0i} & 0.017e^{2.172i} \\ 0.468e^{0.096i} & 0.005e^{1.923i} & 0.017e^{-2.172i} & 0.49e^{-0.0i} \end{pmatrix}$ | | | | 96.2% |
| Reduced $T_2^*$ simulation | $\begin{pmatrix} 0.499e^{0.0i} & 0.003e^{1.687i} & 0.017e^{-2.074i} & 0.405e^{-0.093i} \\ 0.003e^{-1.687i} & 0.009e^{0.0i} & 0.007e^{0.073i} & 0.005e^{-2.411i} \\ 0.017e^{2.074i} & 0.007e^{-0.073i} & 0.01e^{0.0i} & 0.015e^{2.118i} \\ 0.405e^{0.093i} & 0.005e^{2.411i} & 0.015e^{-2.118i} & 0.483e^{0.0i} \end{pmatrix}$ | | | | 89.4% |

**Supplementary Data Table 2 | Experimental vs simulated state tomography for qubits 12**

## Qubits 23

| | Density Matrices | | | | Fidelity |
|---|---|---|---|---|---|
| Raw data | $\begin{pmatrix} 0.105e^{0.0i} & 0.057e^{0.812i} & 0.054e^{0.611i} & 0.029e^{-0.82i} \\ 0.057e^{-0.812i} & 0.425e^{0.0i} & 0.403e^{0.091i} & 0.052e^{0.74i} \\ 0.054e^{-0.611i} & 0.403e^{-0.091i} & 0.449e^{0.0i} & 0.066e^{0.659i} \\ 0.029e^{0.82i} & 0.052e^{-0.74i} & 0.066e^{-0.659i} & 0.021e^{0.0i} \end{pmatrix}$ | | | | 83.8% |
| Processed data | $\begin{pmatrix} 0.075e^{0.0i} & 0.066e^{0.796i} & 0.058e^{0.614i} & 0.02e^{-0.658i} \\ 0.066e^{-0.796i} & 0.439e^{0.0i} & 0.451e^{0.091i} & 0.053e^{0.754i} \\ 0.058e^{-0.614i} & 0.451e^{-0.091i} & 0.471e^{0.0i} & 0.062e^{0.633i} \\ 0.02e^{0.658i} & 0.053e^{-0.754i} & 0.062e^{-0.633i} & 0.016e^{0.0i} \end{pmatrix}$ | | | | 90.4% |
| Full $T_2^*$ simulation | $\begin{pmatrix} 0.01e^{0.0i} & 0.032e^{-1.457i} & 0.032e^{-1.573i} & 0.009e^{-0.095i} \\ 0.032e^{1.457i} & 0.481e^{0.0i} & 0.471e^{0.044i} & 0.017e^{1.246i} \\ 0.032e^{1.573i} & 0.471e^{-0.044i} & 0.5e^{0.0i} & 0.015e^{1.513i} \\ 0.009e^{0.095i} & 0.017e^{-1.246i} & 0.015e^{-1.513i} & 0.008e^{0.0i} \end{pmatrix}$ | | | | 96.1% |
| Reduced $T_2^*$ simulation | $\begin{pmatrix} 0.029e^{0.0i} & 0.027e^{-1.37i} & 0.03e^{-1.944i} & 0.024e^{-0.089i} \\ 0.027e^{1.37i} & 0.463e^{0.0i} & 0.416e^{0.054i} & 0.022e^{0.639i} \\ 0.03e^{1.944i} & 0.416e^{-0.054i} & 0.482e^{0.0i} & 0.012e^{1.514i} \\ 0.024e^{0.089i} & 0.022e^{-0.639i} & 0.012e^{-1.514i} & 0.026e^{0.0i} \end{pmatrix}$ | | | | 88.8% |

**Supplementary Data Table 3 | Experimental vs simulated state tomography for qubits 23**

**Qubits 34**

| | Density Matrices | Fidelity |
|---|---|---|
| Raw data | $\begin{pmatrix} 0.149e^{0.0i} & 0.04e^{1.2i} & 0.055e^{-0.241i} & 0.025e^{1.213i} \\ 0.04e^{-1.2i} & 0.395e^{0.0i} & 0.377e^{-0.173i} & 0.104e^{-1.446i} \\ 0.055e^{0.241i} & 0.377e^{0.173i} & 0.424e^{0.0i} & 0.092e^{-1.393i} \\ 0.025e^{-1.213i} & 0.104e^{1.446i} & 0.092e^{1.393i} & 0.032e^{0.0i} \end{pmatrix}$ | 78.0% |
| Processed data | $\begin{pmatrix} 0.072e^{0.0i} & 0.048e^{0.805i} & 0.055e^{0.08i} & 0.009e^{0.598i} \\ 0.048e^{-0.805i} & 0.434e^{0.0i} & 0.444e^{-0.182i} & 0.105e^{-1.531i} \\ 0.055e^{-0.08i} & 0.444e^{0.182i} & 0.466e^{0.0i} & 0.103e^{-1.389i} \\ 0.009e^{-0.598i} & 0.105e^{1.531i} & 0.103e^{1.389i} & 0.028e^{0.0i} \end{pmatrix}$ | 88.6% |
| Full $T_2^*$ simulation | $\begin{pmatrix} 0.004e^{0.0i} & 0.017e^{-0.771i} & 0.017e^{-1.079i} & 0.003e^{0.014i} \\ 0.017e^{0.771i} & 0.49e^{0.0i} & 0.485e^{-0.264i} & 0.003e^{0.832i} \\ 0.017e^{1.079i} & 0.485e^{0.264i} & 0.503e^{0.0i} & 0.003e^{1.344i} \\ 0.003e^{-0.014i} & 0.003e^{-0.832i} & 0.003e^{-1.344i} & 0.003e^{0.0i} \end{pmatrix}$ | 96.5% |
| Reduced $T_2^*$ simulation | $\begin{pmatrix} 0.009e^{-0.0i} & 0.014e^{-0.723i} & 0.015e^{-1.134i} & 0.007e^{-0.002i} \\ 0.014e^{0.723i} & 0.485e^{0.0i} & 0.429e^{-0.261i} & 0.004e^{0.255i} \\ 0.015e^{1.134i} & 0.429e^{0.261i} & 0.499e^{0.0i} & 0.002e^{0.823i} \\ 0.007e^{0.002i} & 0.004e^{-0.256i} & 0.002e^{-0.827i} & 0.008e^{-0.0i} \end{pmatrix}$ | 90.6% |

**Supplementary Data Table 4 | Experimental vs simulated state tomography for qubits 34**

**Qubits 45**

| | Density Matrices | Fidelity |
|---|---|---|
| Raw data | $\begin{pmatrix} 0.049e^{0.0i} & 0.026e^{-3.07i} & 0.035e^{-2.819i} & 0.012e^{-1.345i} \\ 0.026e^{3.07i} & 0.469e^{0.0i} & 0.444e^{-0.125i} & 0.033e^{-0.174i} \\ 0.035e^{2.819i} & 0.444e^{0.125i} & 0.476e^{0.0i} & 0.037e^{0.005i} \\ 0.012e^{1.345i} & 0.033e^{0.174i} & 0.037e^{-0.005i} & 0.006e^{0.0i} \end{pmatrix}$ | 91.3% |
| Processed data | $\begin{pmatrix} 0.025e^{0.0i} & 0.028e^{-3.057i} & 0.037e^{-2.829i} & 0.006e^{-1.595i} \\ 0.028e^{3.057i} & 0.481e^{0.0i} & 0.481e^{-0.125i} & 0.036e^{-0.181i} \\ 0.037e^{2.829i} & 0.481e^{0.125i} & 0.489e^{0.0i} & 0.038e^{0.027i} \\ 0.006e^{1.595i} & 0.036e^{0.181i} & 0.038e^{-0.027i} & 0.005e^{0.0i} \end{pmatrix}$ | 96.2% |
| Full $T_2^*$ simulation | $\begin{pmatrix} 0.002e^{0.0i} & 0.008e^{0.103i} & 0.007e^{0.12i} & 0.002e^{-0.152i} \\ 0.008e^{-0.103i} & 0.491e^{0.0i} & 0.493e^{-0.005i} & 0.015e^{-1.539i} \\ 0.007e^{-0.12i} & 0.493e^{0.005i} & 0.505e^{0.0i} & 0.016e^{-1.596i} \\ 0.002e^{0.152i} & 0.015e^{1.539i} & 0.016e^{1.596i} & 0.002e^{0.0i} \end{pmatrix}$ | 99.2% |
| Reduced $T_2^*$ simulation | $\begin{pmatrix} 0.009e^{0.0i} & 0.009e^{0.122i} & 0.004e^{0.402i} & 0.009e^{-0.032i} \\ 0.009e^{-0.122i} & 0.483e^{0.0i} & 0.472e^{-0.008i} & 0.016e^{-1.255i} \\ 0.004e^{-0.402i} & 0.472e^{0.008i} & 0.498e^{-0.0i} & 0.015e^{-1.611i} \\ 0.009e^{0.032i} & 0.016e^{1.255i} & 0.015e^{1.611i} & 0.01e^{0.0i} \end{pmatrix}$ | 96.3% |

**Supplementary Data Table 5 | Experimental vs simulated state tomography for qubits 45**

## Qubits 56

| | Density Matrices | Fidelity |
|---|---|---|
| Raw data | $\begin{pmatrix} 0.471e^{0.0i} & 0.031e^{0.392i} & 0.023e^{0.785i} & 0.442e^{-1.471i} \\ 0.031e^{-0.392i} & 0.03e^{0.0i} & 0.014e^{-2.554i} & 0.01e^{-2.85i} \\ 0.023e^{-0.785i} & 0.014e^{2.554i} & 0.024e^{0.0i} & 0.035e^{-1.983i} \\ 0.442e^{1.471i} & 0.01e^{2.85i} & 0.035e^{1.983i} & 0.475e^{0.0i} \end{pmatrix}$ | 91.3% |
| Processed data | $\begin{pmatrix} 0.481e^{0.0i} & 0.032e^{0.322i} & 0.023e^{0.723i} & 0.464e^{-1.472i} \\ 0.032e^{-0.322i} & 0.021e^{0.0i} & 0.012e^{-2.715i} & 0.012e^{-2.715i} \\ 0.023e^{-0.723i} & 0.012e^{2.715i} & 0.012e^{0.0i} & 0.038e^{-1.976i} \\ 0.464e^{1.472i} & 0.012e^{2.715i} & 0.038e^{1.976i} & 0.486e^{0.0i} \end{pmatrix}$ | 94.6% |
| Full $T_2^*$ simulation | $\begin{pmatrix} 0.486e^{0.0i} & 0.022e^{1.851i} & 0.004e^{2.973i} & 0.49e^{-1.525i} \\ 0.022e^{-1.851i} & 0.002e^{0.0i} & 0.001e^{-1.495i} & 0.023e^{2.942i} \\ 0.004e^{-2.973i} & 0.001e^{1.495i} & 0.001e^{0.0i} & 0.004e^{1.595i} \\ 0.49e^{1.525i} & 0.023e^{-2.942i} & 0.004e^{-1.595i} & 0.51e^{0.0i} \end{pmatrix}$ | 98.7% |
| Reduced $T_2^*$ simulation | $\begin{pmatrix} 0.483e^{0.0i} & 0.022e^{1.861i} & 0.005e^{2.481i} & 0.47e^{-1.532i} \\ 0.022e^{-1.861i} & 0.006e^{0.0i} & 0.005e^{-1.526i} & 0.022e^{3.038i} \\ 0.005e^{-2.481i} & 0.005e^{1.526i} & 0.005e^{-0.0i} & 0.004e^{1.576i} \\ 0.47e^{1.532i} & 0.022e^{-3.038i} & 0.004e^{-1.576i} & 0.506e^{0.0i} \end{pmatrix}$ | 96.4% |

**Supplementary Data Table 6 | Experimental vs simulated state tomography for qubits 56**

## Qubits 123

| | Density Matrices | Fidelity |
|---|---|---|
| Raw | (8×8 density matrix) | 64.3% |
| Proc. data | (8×8 density matrix) | 77.0% |
| Full $T_2^*$ sim. | (8×8 density matrix) | 87.9% |
| Red. $T_2^*$ sim. | (8×8 density matrix) | 73.4% |

**Supplementary Data Table 7 | Experimental vs simulated state tomography for qubits 123**

13

## Qubits 234

| | Density Matrices | Fidelity |
|---|---|---|
| Raw | $\begin{pmatrix} 0.455e^{0.0i} & 0.04e^{0.306i} & 0.025e^{-1.365i} & 0.007e^{-1.571i} & 0.117e^{1.214i} & 0.014e^{-2.554i} & 0.031e^{2.492i} & 0.168e^{0.295i} \\ 0.04e^{-0.306i} & 0.04e^{-0.0i} & 0.01e^{3.042i} & 0.003e^{1.249i} & 0.004e^{-0.0i} & 0.003e^{0.785i} & 0.043e^{2.489i} & 0.01e^{-0.197i} \\ 0.025e^{1.365i} & 0.01e^{-3.042i} & 0.027e^{-0.0i} & 0.007e^{2.863i} & 0.007e^{0.464i} & 0.021e^{2.253i} & 0.007e^{0.983i} & 0.021e^{1.759i} \\ 0.007e^{1.571i} & 0.003e^{-1.249i} & 0.007e^{-2.863i} & 0.033e^{0.0i} & 0.014e^{-3.142i} & 0.017e^{-1.688i} & 0.002e^{1.571i} & 0.024e^{-0.165i} \\ 0.117e^{-1.214i} & 0.004e^{0.0i} & 0.007e^{-0.464i} & 0.014e^{3.142i} & 0.07e^{0.0i} & 0.021e^{2.322i} & 0.027e^{0.993i} & 0.075e^{-2.271i} \\ 0.014e^{2.554i} & 0.003e^{-0.785i} & 0.021e^{-2.253i} & 0.017e^{1.688i} & 0.021e^{-2.322i} & 0.031e^{-0.0i} & 0.012e^{-0.427i} & 0.02e^{1.153i} \\ 0.031e^{-2.492i} & 0.043e^{-2.489i} & 0.007e^{-0.983i} & 0.002e^{-1.571i} & 0.027e^{-0.993i} & 0.012e^{0.427i} & 0.08e^{0.0i} & 0.068e^{2.843i} \\ 0.168e^{-0.295i} & 0.01e^{0.197i} & 0.021e^{-1.759i} & 0.024e^{0.165i} & 0.075e^{2.271i} & 0.02e^{-1.153i} & 0.068e^{-2.843i} & 0.265e^{0.0i} \end{pmatrix}$ | 52.8% |
| Proc. data | $\begin{pmatrix} 0.559e^{0.0i} & 0.051e^{0.236i} & 0.034e^{-1.268i} & 0.008e^{-0.245i} & 0.139e^{1.361i} & 0.021e^{-2.459i} & 0.058e^{2.693i} & 0.288e^{0.256i} \\ 0.051e^{-0.236i} & 0.011e^{-0.0i} & 0.003e^{-1.893i} & 0.001e^{-0.0i} & 0.018e^{1.18i} & 0.004e^{-2.678i} & 0.019e^{2.393i} & 0.032e^{-0.492i} \\ 0.034e^{1.268i} & 0.003e^{1.893i} & 0.003e^{0.0i} & 0.002e^{2.678i} & 0.004e^{3.142i} & 0.001e^{0.785i} & 0.005e^{-0.927i} & 0.027e^{1.916i} \\ 0.008e^{0.245i} & 0.001e^{0.0i} & 0.002e^{-2.678i} & 0.005e^{-0.0i} & 0.013e^{2.642i} & 0.004e^{-1.326i} & 0.008e^{-2.897i} & 0.028e^{-0.071i} \\ 0.139e^{-1.361i} & 0.018e^{-1.18i} & 0.004e^{-3.142i} & 0.013e^{-2.642i} & 0.068e^{-0.0i} & 0.016e^{2.313i} & 0.042e^{1.022i} & 0.089e^{-1.999i} \\ 0.021e^{2.459i} & 0.004e^{2.678i} & 0.001e^{-0.785i} & 0.004e^{1.326i} & 0.016e^{-2.313i} & 0.004e^{0.0i} & 0.012e^{-1.326i} & 0.026e^{1.532i} \\ 0.058e^{-2.693i} & 0.019e^{-2.393i} & 0.005e^{0.927i} & 0.008e^{2.897i} & 0.042e^{-1.022i} & 0.012e^{1.326i} & 0.045e^{0.0i} & 0.08e^{2.903i} \\ 0.288e^{-0.256i} & 0.032e^{0.492i} & 0.027e^{-1.916i} & 0.028e^{0.071i} & 0.089e^{1.999i} & 0.026e^{-1.532i} & 0.08e^{-2.903i} & 0.305e^{0.0i} \end{pmatrix}$ | 72.0% |
| Full $T_2^*$ sim. | $\begin{pmatrix} 0.501e^{0.0i} & 0.002e^{-2.311i} & 0.0e^{2.872i} & 0.012e^{-2.596i} & 0.009e^{-0.058i} & 0.001e^{1.784i} & 0.011e^{0.992i} & 0.401e^{-1.621i} \\ 0.002e^{2.311i} & 0.003e^{0.0i} & 0.0e^{0.761i} & 0.0e^{-2.899i} & 0.0e^{-2.047i} & 0.0e^{-0.212i} & 0.003e^{1.792i} & 0.001e^{1.399i} \\ 0.0e^{-2.872i} & 0.0e^{-0.761i} & 0.0e^{0.0i} & 0.0e^{-2.742i} & 0.0e^{-1.45i} & 0.0e^{-1.831i} & 0.0e^{1.549i} & 0.0e^{1.456i} \\ 0.012e^{2.596i} & 0.0e^{2.899i} & 0.0e^{2.742i} & 0.002e^{-0.0i} & 0.001e^{1.541i} & 0.0e^{-1.561i} & 0.0e^{-1.99i} & 0.009e^{1.532i} \\ 0.009e^{0.058i} & 0.0e^{2.047i} & 0.0e^{1.45i} & 0.001e^{-1.541i} & 0.002e^{-0.0i} & 0.0e^{1.91i} & 0.0e^{1.44i} & 0.005e^{-1.527i} \\ 0.001e^{-1.784i} & 0.0e^{0.212i} & 0.0e^{1.831i} & 0.0e^{1.561i} & 0.0e^{-1.91i} & 0.0e^{0.0i} & 0.0e^{1.596i} & 0.001e^{2.882i} \\ 0.011e^{-0.992i} & 0.003e^{-1.792i} & 0.0e^{-1.549i} & 0.0e^{1.99i} & 0.0e^{-1.44i} & 0.0e^{-1.596i} & 0.004e^{0.0i} & 0.013e^{-2.669i} \\ 0.401e^{1.621i} & 0.001e^{-1.399i} & 0.0e^{-1.456i} & 0.009e^{-1.532i} & 0.005e^{1.527i} & 0.001e^{-2.882i} & 0.013e^{2.669i} & 0.489e^{0.0i} \end{pmatrix}$ | 89.6% |
| Red. $T_2^*$ sim. | $\begin{pmatrix} 0.486e^{0.0i} & 0.008e^{-1.944i} & 0.0e^{2.517i} & 0.018e^{-2.075i} & 0.007e^{0.007i} & 0.0e^{1.835i} & 0.008e^{0.797i} & 0.244e^{-1.651i} \\ 0.008e^{1.944i} & 0.017e^{-0.0i} & 0.001e^{1.218i} & 0.0e^{0.0i} & 0.0e^{2.071i} & 0.0e^{0.003i} & 0.009e^{1.745i} & 0.001e^{2.349i} \\ 0.0e^{-2.517i} & 0.001e^{-1.218i} & 0.0e^{-0.0i} & 0.0e^{3.039i} & 0.0e^{-1.004i} & 0.0e^{-1.867i} & 0.0e^{1.76i} & 0.0e^{1.609i} \\ 0.018e^{2.075i} & 0.0e^{-1.088i} & 0.0e^{-3.039i} & 0.004e^{-0.0i} & 0.002e^{1.412i} & 0.0e^{-2.537i} & 0.0e^{-2.224i} & 0.005e^{1.636i} \\ 0.007e^{-0.007i} & 0.0e^{-2.071i} & 0.0e^{1.004i} & 0.002e^{-1.412i} & 0.004e^{-0.0i} & 0.0e^{-2.041i} & 0.0e^{-2.251i} & 0.005e^{1.475i} \\ 0.0e^{-1.835i} & 0.0e^{-0.003i} & 0.0e^{1.867i} & 0.0e^{2.537i} & 0.0e^{2.041i} & 0.0e^{-0.0i} & 0.0e^{-1.677i} & 0.001e^{3.008i} \\ 0.008e^{-0.797i} & 0.009e^{-1.745i} & 0.0e^{-1.76i} & 0.0e^{2.224i} & 0.0e^{2.251i} & 0.0e^{1.677i} & 0.017e^{0.0i} & 0.009e^{-3.073i} \\ 0.244e^{1.651i} & 0.001e^{-2.349i} & 0.0e^{-1.609i} & 0.005e^{-1.636i} & 0.005e^{-1.475i} & 0.001e^{-3.008i} & 0.009e^{3.073i} & 0.471e^{0.0i} \end{pmatrix}$ | 72.2% |

**Supplementary Data Table 8 | Experimental vs simulated state tomography for qubits 234**

## Qubits 345

| | Density Matrices | Fidelity |
|---|---|---|
| Raw | $\begin{pmatrix} 0.419e^{-0.0i} & 0.066e^{-3.036i} & 0.021e^{1.326i} & 0.032e^{-0.351i} & 0.055e^{1.278i} & 0.014e^{2.2i} & 0.016e^{-1.695i} & 0.163e^{-1.626i} \\ 0.066e^{3.036i} & 0.045e^{-0.0i} & 0.011e^{-1.107i} & 0.003e^{0.785i} & 0.006e^{2.111i} & 0.004e^{0.245i} & 0.029e^{2.971i} & 0.034e^{1.541i} \\ 0.021e^{-1.326i} & 0.011e^{-1.107i} & 0.038e^{-0.0i} & 0.009e^{-2.034i} & 0.013e^{0.322i} & 0.045e^{-0.223i} & 0.006e^{2.246i} & 0.013e^{-0.733i} \\ 0.032e^{0.351i} & 0.003e^{-0.785i} & 0.009e^{2.034i} & 0.03e^{-0.0i} & 0.029e^{0.51i} & 0.018e^{2.476i} & 0.008e^{-1.166i} & 0.009e^{-0.709i} \\ 0.055e^{-1.278i} & 0.006e^{-2.111i} & 0.013e^{-0.322i} & 0.029e^{-0.51i} & 0.053e^{-0.0i} & 0.006e^{2.111i} & 0.006e^{2.467i} & 0.063e^{-2.492i} \\ 0.014e^{-2.2i} & 0.004e^{-0.245i} & 0.045e^{0.223i} & 0.018e^{-2.476i} & 0.006e^{-2.111i} & 0.064e^{-0.0i} & 0.001e^{0.785i} & 0.02e^{0.785i} \\ 0.016e^{1.695i} & 0.029e^{-2.971i} & 0.006e^{-2.246i} & 0.008e^{1.166i} & 0.006e^{-2.467i} & 0.001e^{-0.785i} & 0.044e^{-0.0i} & 0.055e^{0.862i} \\ 0.163e^{1.626i} & 0.034e^{-1.541i} & 0.013e^{0.733i} & 0.009e^{0.709i} & 0.063e^{2.492i} & 0.02e^{-0.785i} & 0.055e^{-0.862i} & 0.308e^{-0.0i} \end{pmatrix}$ | 52.7% |
| Proc. data | $\begin{pmatrix} 0.508e^{0.0i} & 0.088e^{-3.051i} & 0.025e^{-1.373i} & 0.051e^{-0.159i} & 0.084e^{1.139i} & 0.029e^{-2.207i} & 0.032e^{-1.633i} & 0.295e^{-1.618i} \\ 0.088e^{3.051i} & 0.021e^{0.0i} & 0.004e^{0.983i} & 0.006e^{-2.976i} & 0.012e^{-1.92i} & 0.008e^{0.519i} & 0.008e^{2.356i} & 0.047e^{1.528i} \\ 0.025e^{1.373i} & 0.004e^{-0.983i} & 0.023e^{-0.0i} & 0.011e^{-1.951i} & 0.006e^{0.322i} & 0.026e^{-0.19i} & 0.006e^{-2.82i} & 0.021e^{-0.82i} \\ 0.051e^{0.159i} & 0.006e^{2.976i} & 0.011e^{1.951i} & 0.009e^{0.0i} & 0.009e^{1.46i} & 0.013e^{1.966i} & 0.006e^{-1.406i} & 0.021e^{-1.279i} \\ 0.084e^{-1.139i} & 0.012e^{1.92i} & 0.006e^{-0.322i} & 0.009e^{-1.46i} & 0.023e^{-0.0i} & 0.002e^{1.107i} & 0.018e^{3.031i} & 0.085e^{-2.573i} \\ 0.029e^{2.207i} & 0.008e^{-0.519i} & 0.026e^{0.19i} & 0.013e^{-1.966i} & 0.002e^{-1.107i} & 0.033e^{0.0i} & 0.002e^{3.142i} & 0.028e^{0.608i} \\ 0.032e^{1.633i} & 0.008e^{-2.356i} & 0.006e^{2.82i} & 0.006e^{1.406i} & 0.018e^{-3.031i} & 0.002e^{-3.142i} & 0.023e^{0.0i} & 0.067e^{-0.828i} \\ 0.295e^{1.618i} & 0.047e^{-1.528i} & 0.021e^{0.82i} & 0.021e^{1.279i} & 0.085e^{2.573i} & 0.028e^{-0.608i} & 0.067e^{0.828i} & 0.361e^{0.0i} \end{pmatrix}$ | 72.9% |
| Full $T_2^*$ sim. | $\begin{pmatrix} 0.505e^{-0.0i} & 0.01e^{-3.135i} & 0.0e^{-1.611i} & 0.017e^{-2.681i} & 0.009e^{-0.231i} & 0.0e^{1.195i} & 0.009e^{1.029i} & 0.436e^{-1.593i} \\ 0.01e^{3.135i} & 0.001e^{-0.0i} & 0.0e^{0.719i} & 0.0e^{0.599i} & 0.0e^{3.079i} & 0.0e^{-0.815i} & 0.001e^{1.851i} & 0.009e^{1.588i} \\ 0.0e^{1.611i} & 0.0e^{-0.719i} & 0.0e^{0.0i} & 0.0e^{-2.441i} & 0.0e^{-0.707i} & 0.0e^{-1.69i} & 0.0e^{1.904i} & 0.0e^{0.198i} \\ 0.017e^{2.681i} & 0.0e^{-0.599i} & 0.0e^{2.441i} & 0.004e^{0.0i} & 0.003e^{1.874i} & 0.0e^{-1.456i} & 0.0e^{-1.86i} & 0.013e^{1.704i} \\ 0.009e^{0.231i} & 0.0e^{-3.079i} & 0.0e^{0.707i} & 0.003e^{-1.874i} & 0.003e^{-0.0i} & 0.0e^{3.088i} & 0.0e^{2.633i} & 0.004e^{-0.028i} \\ 0.0e^{-1.195i} & 0.0e^{0.815i} & 0.0e^{1.69i} & 0.0e^{1.456i} & 0.0e^{-3.088i} & 0.0e^{0.0i} & 0.0e^{-2.822i} & 0.001e^{-2.469i} \\ 0.009e^{-1.029i} & 0.001e^{-1.851i} & 0.0e^{-1.904i} & 0.0e^{1.86i} & 0.0e^{-2.633i} & 0.0e^{2.822i} & 0.001e^{0.0i} & 0.01e^{-2.652i} \\ 0.436e^{1.593i} & 0.009e^{-1.588i} & 0.0e^{-0.198i} & 0.013e^{-1.704i} & 0.004e^{0.028i} & 0.001e^{2.469i} & 0.01e^{2.652i} & 0.485e^{0.0i} \end{pmatrix}$ | 93.1% |
| Red. $T_2^*$ sim. | $\begin{pmatrix} 0.487e^{-0.0i} & 0.011e^{-2.918i} & 0.001e^{0.452i} & 0.046e^{-1.967i} & 0.008e^{-0.127i} & 0.0e^{1.278i} & 0.006e^{0.929i} & 0.306e^{-1.619i} \\ 0.011e^{2.918i} & 0.005e^{0.0i} & 0.0e^{1.38i} & 0.001e^{1.284i} & 0.0e^{-3.015i} & 0.0e^{-0.068i} & 0.003e^{1.696i} & 0.006e^{1.584i} \\ 0.001e^{-0.452i} & 0.0e^{-1.38i} & 0.0e^{-0.0i} & 0.0e^{-2.864i} & 0.0e^{-0.839i} & 0.0e^{-1.559i} & 0.0e^{1.747i} & 0.0e^{0.092i} \\ 0.046e^{1.967i} & 0.001e^{-1.284i} & 0.0e^{2.864i} & 0.018e^{-0.0i} & 0.008e^{1.817i} & 0.0e^{-1.418i} & 0.0e^{-1.828i} & 0.01e^{1.739i} \\ 0.008e^{0.127i} & 0.0e^{3.015i} & 0.0e^{0.839i} & 0.008e^{-1.817i} & 0.017e^{-0.0i} & 0.0e^{-2.874i} & 0.001e^{-2.46i} & 0.035e^{1.259i} \\ 0.0e^{-1.278i} & 0.0e^{0.068i} & 0.0e^{1.559i} & 0.0e^{1.418i} & 0.0e^{2.874i} & 0.0e^{-0.0i} & 0.0e^{-1.608i} & 0.001e^{-2.041i} \\ 0.006e^{-0.929i} & 0.003e^{-1.696i} & 0.0e^{-1.747i} & 0.0e^{1.828i} & 0.001e^{2.46i} & 0.0e^{1.608i} & 0.004e^{0.0i} & 0.008e^{-2.818i} \\ 0.306e^{1.619i} & 0.006e^{-1.584i} & 0.0e^{-0.092i} & 0.01e^{-1.739i} & 0.035e^{-1.259i} & 0.001e^{2.041i} & 0.008e^{2.818i} & 0.468e^{0.0i} \end{pmatrix}$ | 78.4% |

**Supplementary Data Table 9 | Experimental vs simulated state tomography for qubits 345**

**Qubits 456**

| | Density Matrices | Fidelity |
|---|---|---|
| Raw | $\begin{pmatrix} 0.425e^{-0.0i} & 0.026e^{0.951i} & 0.038e^{2.246i} & 0.013e^{0.838i} & 0.013e^{-0.0i} & 0.008e^{-0.785i} & 0.064e^{-2.09i} & 0.298e^{-2.702i} \\ 0.026e^{-0.951i} & 0.025e^{-0.0i} & 0.015e^{0.133i} & 0.026e^{2.575i} & 0.006e^{-2.82i} & 0.004e^{0.983i} & 0.006e^{3.142i} & 0.025e^{3.062i} \\ 0.038e^{-2.246i} & 0.015e^{-0.133i} & 0.062e^{-0.0i} & 0.017e^{1.34i} & 0.013e^{-1.249i} & 0.024e^{1.785i} & 0.008e^{-1.052i} & 0.012e^{1.654i} \\ 0.013e^{-0.838i} & 0.026e^{-2.575i} & 0.017e^{-1.34i} & 0.053e^{0.0i} & 0.001e^{0.785i} & 0.005e^{1.373i} & 0.011e^{-1.661i} & 0.025e^{0.373i} \\ 0.013e^{0.0i} & 0.006e^{2.82i} & 0.013e^{1.249i} & 0.001e^{-0.785i} & 0.051e^{-0.0i} & 0.018e^{3.031i} & 0.018e^{1.107i} & 0.007e^{2.034i} \\ 0.008e^{0.785i} & 0.004e^{-0.983i} & 0.024e^{-1.785i} & 0.005e^{-1.373i} & 0.018e^{-3.031i} & 0.018e^{-0.0i} & 0.009e^{-1.46i} & 0.012e^{2.897i} \\ 0.064e^{2.09i} & 0.006e^{-3.142i} & 0.008e^{1.052i} & 0.011e^{1.661i} & 0.018e^{-1.107i} & 0.009e^{1.46i} & 0.043e^{-0.0i} & 0.043e^{0.07i} \\ 0.298e^{2.702i} & 0.025e^{-3.062i} & 0.012e^{-1.654i} & 0.025e^{-0.373i} & 0.007e^{-2.034i} & 0.012e^{-2.897i} & 0.043e^{-0.07i} & 0.323e^{-0.0i} \end{pmatrix}$ | 67.2% |
| Proc. data | $\begin{pmatrix} 0.477e^{-0.0i} & 0.012e^{1.03i} & 0.042e^{2.339i} & 0.012e^{0.54i} & 0.005e^{-1.373i} & 0.009e^{-0.464i} & 0.07e^{-2.06i} & 0.398e^{-2.7i} \\ 0.012e^{-1.03i} & 0.008e^{-0.0i} & 0.012e^{-0.245i} & 0.008e^{2.897i} & 0.001e^{-2.356i} & 0.006e^{1.249i} & 0.001e^{1.571i} & 0.026e^{3.027i} \\ 0.042e^{-2.339i} & 0.012e^{0.245i} & 0.037e^{-0.0i} & 0.008e^{3.017i} & 0.016e^{-1.695i} & 0.027e^{1.798i} & 0.006e^{0.785i} & 0.022e^{1.849i} \\ 0.012e^{-0.54i} & 0.008e^{-2.897i} & 0.008e^{-3.017i} & 0.013e^{-0.0i} & 0.004e^{-2.356i} & 0.002e^{0.0i} & 0.007e^{-2.034i} & 0.016e^{0.245i} \\ 0.005e^{1.373i} & 0.001e^{2.356i} & 0.016e^{1.695i} & 0.004e^{2.356i} & 0.014e^{-0.0i} & 0.018e^{-2.856i} & 0.007e^{0.785i} & 0.001e^{2.356i} \\ 0.009e^{0.464i} & 0.006e^{-1.249i} & 0.027e^{-1.798i} & 0.002e^{-0.0i} & 0.018e^{2.856i} & 0.027e^{-0.0i} & 0.009e^{-2.191i} & 0.012e^{-3.058i} \\ 0.07e^{2.06i} & 0.001e^{-1.571i} & 0.006e^{-0.785i} & 0.007e^{2.034i} & 0.007e^{-0.785i} & 0.009e^{2.191i} & 0.016e^{-0.0i} & 0.048e^{-0.501i} \\ 0.398e^{2.7i} & 0.026e^{-3.027i} & 0.022e^{-1.849i} & 0.016e^{-0.245i} & 0.001e^{-2.356i} & 0.012e^{3.058i} & 0.048e^{0.501i} & 0.408e^{-0.0i} \end{pmatrix}$ | 84.1% |
| Full T$_2^*$ sim. | $\begin{pmatrix} 0.506e^{-0.0i} & 0.02e^{-3.108i} & 0.0e^{-1.711i} & 0.009e^{-2.784i} & 0.011e^{1.122i} & 0.0e^{-2.141i} & 0.011e^{-0.069i} & 0.46e^{-1.496i} \\ 0.02e^{3.108i} & 0.002e^{-0.0i} & 0.0e^{0.721i} & 0.0e^{0.354i} & 0.0e^{-2.052i} & 0.0e^{1.072i} & 0.001e^{2.121i} & 0.018e^{1.63i} \\ 0.0e^{1.711i} & 0.0e^{-0.721i} & 0.0e^{-0.0i} & 0.0e^{-1.364i} & 0.0e^{0.185i} & 0.0e^{1.674i} & 0.0e^{3.047i} & 0.0e^{0.532i} \\ 0.009e^{2.784i} & 0.0e^{-0.354i} & 0.0e^{1.364i} & 0.001e^{0.0i} & 0.001e^{1.674i} & 0.0e^{-1.443i} & 0.0e^{3.047i} & 0.008e^{1.578i} \\ 0.011e^{-1.122i} & 0.0e^{2.052i} & 0.0e^{-0.185i} & 0.001e^{-1.674i} & 0.001e^{0.0i} & 0.0e^{-3.131i} & 0.0e^{-1.442i} & 0.011e^{-2.825i} \\ 0.0e^{2.141i} & 0.0e^{-1.072i} & 0.0e^{1.978i} & 0.0e^{1.443i} & 0.0e^{3.131i} & 0.0e^{-0.0i} & 0.0e^{0.706i} & 0.0e^{0.43i} \\ 0.011e^{0.069i} & 0.001e^{-2.121i} & 0.0e^{-1.674i} & 0.0e^{-3.047i} & 0.0e^{1.442i} & 0.0e^{-0.706i} & 0.001e^{-0.0i} & 0.012e^{-1.417i} \\ 0.46e^{1.496i} & 0.018e^{-1.63i} & 0.0e^{-0.532i} & 0.008e^{-1.578i} & 0.011e^{2.825i} & 0.0e^{-0.43i} & 0.012e^{1.417i} & 0.488e^{-0.0i} \end{pmatrix}$ | 95.7% |
| Red. T$_2^*$ sim. | $\begin{pmatrix} 0.502e^{0.0i} & 0.019e^{-3.07i} & 0.0e^{-1.149i} & 0.014e^{-2.15i} & 0.008e^{1.123i} & 0.0e^{-2.499i} & 0.008e^{0.003i} & 0.322e^{-1.485i} \\ 0.019e^{3.07i} & 0.003e^{0.0i} & 0.0e^{1.061i} & 0.001e^{0.812i} & 0.0e^{-2.372i} & 0.0e^{1.046i} & 0.002e^{1.814i} & 0.012e^{1.648i} \\ 0.0e^{1.149i} & 0.0e^{-1.061i} & 0.0e^{-0.0i} & 0.0e^{-1.547i} & 0.0e^{-0.05i} & 0.0e^{-1.724i} & 0.0e^{1.464i} & 0.0e^{0.92i} \\ 0.014e^{2.15i} & 0.001e^{-0.812i} & 0.0e^{1.547i} & 0.005e^{-0.0i} & 0.003e^{1.592i} & 0.0e^{-1.601i} & 0.0e^{2.5i} & 0.006e^{1.544i} \\ 0.008e^{-1.123i} & 0.0e^{2.372i} & 0.0e^{0.05i} & 0.003e^{-1.592i} & 0.005e^{-0.0i} & 0.0e^{-3.067i} & 0.0e^{-1.701i} & 0.011e^{2.717i} \\ 0.0e^{2.499i} & 0.0e^{-1.046i} & 0.0e^{1.724i} & 0.0e^{1.601i} & 0.0e^{3.067i} & 0.0e^{-0.0i} & 0.0e^{-0.312i} & 0.0e^{-0.553i} \\ 0.008e^{-0.003i} & 0.002e^{-1.814i} & 0.0e^{-1.464i} & 0.0e^{-2.5i} & 0.0e^{1.701i} & 0.0e^{0.312i} & 0.002e^{0.0i} & 0.011e^{-1.472i} \\ 0.322e^{1.485i} & 0.012e^{-1.648i} & 0.0e^{-0.92i} & 0.006e^{-1.544i} & 0.011e^{-2.717i} & 0.0e^{0.553i} & 0.011e^{1.472i} & 0.483e^{0.0i} \end{pmatrix}$ | 81.4% |

**Supplementary Data Table 10 | Experimental vs simulated state tomography for qubits 456**

	
\end{document}